\immediate\write18{makeindex \jobname.nlo -s nomencl.ist -o \jobname.nls}
\documentclass[preprint,12pt]{elsarticle}
\usepackage[utf8]{inputenc}
\usepackage[colorlinks=true,linkcolor=blue,urlcolor=blue,citecolor=blue,
anchorcolor = blue]{hyperref}

\usepackage{graphicx,subcaption}
\setkeys{Gin}{keepaspectratio}
\usepackage{booktabs,multirow,multicol,makecell}
\usepackage{floatrow}
\DeclareFloatFont{scriptsize}{\scriptsize}
\usepackage{svg}
\usepackage{amssymb}
\usepackage{amsmath}
\usepackage{cleveref}
\usepackage{lineno}
\usepackage{mathtools}
\usepackage{enumitem}
\usepackage{endfloat}

\biboptions{semicolon,square,sort&compress}

\makeatletter
\def\ps@pprintTitle{%
	\let\@oddhead\@empty
	\let\@evenhead\@empty
	\def\@oddfoot{\centerline{\thepage}}%
	\let\@evenfoot\@oddfoot}
\makeatother

\usepackage{nomencl}
\usepackage{ifthen}
\usepackage{etoolbox}
\renewcommand{\nomgroup}[1]{%
	\item[\footnotesize\bfseries
	\ifstrequal{#1}{A}{Variables}{%
	\ifstrequal{#1}{B}{Constants \& Fixed Parameters}{%
	\ifstrequal{#1}{C}{Uncertainty Quantification}{%
	\ifstrequal{#1}{D}{Subscripts}{}}}}
]}
\makenomenclature

\begin{document}

\begin{frontmatter}

\title{Development of an aqueous lignin mixture thermophysical model for hydrothermal liquefaction applications using uncertainty quantification tools}

%%------------AUTHORS--------------
\author[ept]{A. Dinis S. Nunes}\ead{adnunes@ntnu.no}
\author[uva]{José Sierra-Pallares}\ead{jsierra@eii.uva.es}
\author[ept]{Khanh-Quang Tran}\ead{khanh-quang.tran@ntnu.no}
\author[ept]{R. Jason Hearst\corref{cor1}}
\cortext[cor1]{Corresponding author.\ead{jason.hearst@ntnu.no} Tel.: +4791359427}
%%------------/AUTHORS--------------

%%------------ADDRESSES--------------
\address[ept]{Department of Energy and Processing Engineering, Norwegian University of Science and Technology}
\address[uva]{Department of Energy Engineering and Fluid Mechanics, University of Valladolid}
%%------------/ADDRESSES--------------

\begin{abstract}\label{sec:abstract}
	A thermophysical model is developed that can predict the properties of two lignin mixtures, black liquor and lignosulfonates, up to 50\% mass fractions, at hydrothermal conditions. An uncertainty quantification framework linked with classic thermodynamical modelling was included to account for the extreme variability of the raw material. An idealised flow simulation verified the model, where hot compressed water mixes with a cold, aqueous lignin stream in a T-piece reactor configuration. The uncertainty quantification procedure determined that density and heat capacity uncertainty significantly influence residence time, and viscosity uncertainty mainly affects mixing. Micromixing time is five-fold and ten-fold higher for black liquor and lignosulfonates mixtures, respectively, compared to pure water mixing. The uncertainty in all simulated quantities of interest caused by the thermophysical model is reduced by increasing flow rates. This study predicted chemical reactor behaviour under varying thermophysical conditions and their final effect in terms of confidence intervals.
\end{abstract}


\end{frontmatter}

\nomenclature[A,00]{$T$}{Temperature}
\nomenclature[A,01]{$\phi$}{Volume fraction}
\nomenclature[A,02]{$\varphi$}{Mass fraction}
\nomenclature[A,1]{$\rho$}{Density}
\nomenclature[A,20]{$C_P$}{Heat capacity}
\nomenclature[A,21]{$\Phi_{C_P}$}{Heat capacity polynomial}
\nomenclature[A,300]{$\eta$}{Dynamic viscosity}
\nomenclature[A,301]{$\nu$}{Kinematic viscosity}
\nomenclature[A,31]{K}{Consistency index}
\nomenclature[A,32]{n}{Power law index}
\nomenclature[A,33]{$\dot{\gamma}$}{Shear rate}
\nomenclature[A,340]{$\eta_0$}{Zero-shear viscosity}
\nomenclature[A,341]{$\xi$}{Plug size}
%\nomenclature[A,342]{$R_c$}{Plug radius}
\nomenclature[A,343]{$\phi_\mathrm{m}$}{Maximum packing fraction}
\nomenclature[A,3440]{$\phi_\mathrm{c}$}{Critical volume fraction}
%\nomenclature[A,3441]{$\left\langle v_{e x}\right\rangle$}{Average excluded volume}
%\nomenclature[A,3442]{$\left\langle V_{ex}\right\rangle$ }{Critical total average excluded volume}
%\nomenclature[A,3443]{$V$}{Single particle volume}
\nomenclature[A,3444]{$\varepsilon$}{Particle eccentricity}
\nomenclature[A,3445]{$A_{\mathrm{p}}$}{Particle aspect ratio}
\nomenclature[A,345]{$\left[\eta\right]$}{Intrinsic viscosity}
\nomenclature[A,40]{$\mathcal{D}$}{Molecular diffusivity}
\nomenclature[A,41]{$\mathcal{D}_{\text{mix}}$}{Effective diffusion}
\nomenclature[A,42]{$\varepsilon_\mathcal{D}$}{Virtual diffusion}
%\nomenclature[A,43]{$\chi$}{Fluid resistance}
\nomenclature[A,50]{$Re$}{Reynolds number}
\nomenclature[A,51]{$U$}{Fluid mean velocity}
\nomenclature[A,52]{$c_\text{tr}$}{Tracer concentration}
\nomenclature[A,53]{$\tau_m$}{Mixing time}
%\nomenclature[A,531]{$\epsilon$}{Turbulent kinetic energy mean dissipation rate}
\nomenclature[A,54]{$T_\text{mix}$}{Frozen adiabatic mixing temperature}

\nomenclature[B,10]{$C_{h,0}$}{Heat capacity polynomial zero order constant}
\nomenclature[B,11]{$C_{h,1}$}{Heat capacity polynomial first order constant}
\nomenclature[B,20]{$A$}{Zero-shear viscosity pre-exponential factor}
%\nomenclature[B,210]{$B$}{Activation energy}
\nomenclature[B,211]{$B_1$}{Activation energy polynomial zero order constant}
\nomenclature[B,212]{$B_2$}{Activation energy polynomial first order constant}
%\nomenclature[B,220]{$T_0$}{Glass transition temperature}
\nomenclature[B,221]{$C$}{Glass transition temperature constant}
%\nomenclature[B,230]{$K_{MH}$}{Mark-Houwink eq. constant}
%\nomenclature[B,231]{$\alpha_{MH}$}{Mark-Houwink eq. exponent}
\nomenclature[B,240]{$k_B$}{Boltzmann constant}
\nomenclature[B,241]{$r_p$}{Particle radius}
%\nomenclature[B,242]{$m$}{Particle radius eq. constant}
%\nomenclature[B,243]{$p$}{Particle radius eq. exponent}
\nomenclature[B,30]{$c_\text{tr,in}$}{Inlet tracer concentration}
	
%\nomenclature[C,10]{$\mathcal{K}_{()}$}{Uncertain parameter}
\nomenclature[C,11]{$\mathcal{K}_{\rho}$}{Lignin density}
\nomenclature[C,12]{$\mathcal{K}_{f_H}$}{Heat capacity multiplicative factor}
\nomenclature[C,13]{$\mathcal{K}_{f_B}$}{Zero-shear viscosity multiplicative factor}
\nomenclature[C,14]{$\mathcal{K}_{\beta_2}$}{Plug size exponent}
\nomenclature[C,15]{$\mathcal{K}_{M_W}$}{Lignin molecular weight}
\nomenclature[C,150]{$R$}{Pipe radius}
%\nomenclature[C,151]{$\phi_{\mathrm{m}_1}$}{Perfect spheres maximum packing fraction}
%\nomenclature[C,152]{$b$}{Perfect sphere deviation factor}
%\nomenclature[C,20]{$\boldsymbol{x}$}{Independent variable set}
%\nomenclature[C,21]{$\boldsymbol{\mathcal{K}}$}{Input parameters set}
%\nomenclature[C,30]{$Y$}{Model output}
%\nomenclature[C,31]{$\mathbb{E}$}{Model output mean}
%\nomenclature[C,32]{$\mathbb{V}$}{Model output variance}
\nomenclature[C,33]{$S_{()}$}{1st-order Sobol indices}
%\nomenclature[C,40]{$P_x$}{Percentiles}
%\nomenclature[C,41]{$I_x$}{Prediction interval}

\nomenclature[D,1]{mix}{Mixture}
\nomenclature[D,2]{w}{Water}
\nomenclature[D,3]{s}{Lignin}
\nomenclature[D,4]{c}{Cold stream}
\nomenclature[D,5]{h}{Hot stream}
\nomenclature[D,6]{tr}{Tracer}
\nomenclature[D,7]{tot}{Total mass flow}

\begin{multicols}{2}
	\printnomenclature
\end{multicols}

%\linenumbers %% To activate line numbering

%%------------MAIN BODY--------------
\section{Introduction}\label{sec:intro}

Hydrothermal liquefaction (HTL) is a thermal degradation process that can convert wet biomass substrates into a mineral crude oil analogue referred to as biocrude or bio-oil. The HTL process is carried out at temperatures between $250-374 ^{\circ}\mathrm{C}$ and pressures starting at 4 up to 22 MPa \citep{Elliott_2015}. The solvent and reaction medium used is usually pure water, translating into low environmental impact \citep{Cao_2017}. The processing conditions promote ionic reactions, while organic compounds become soluble in water due to their lower dielectric constant \citep{Lee_2012}. Consequently, oil products are favoured over coke and gases, both formed by radical reactions \citep{Castello_2018}. Also, using hot compressed water minimises corrosion and inorganic precipitation compared to supercritical water \citep{Peterson_2008}. Since water acts as a solvent in HTL, wet biomass substrates can be used directly without the need for feedstock drying, with potential energy savings ranging from 0.3 to 1.66 GJ/(t$\cdot$h) \citep{Haque_2013}. The main HTL product, biocrude, has higher energy density and lower oxygen content than its raw feedstock and can be further processed into biofuels and bioproducts \citep{Elliott_2015}.

This study focuses solely on aqueous lignin biomass feedstocks, particularly black liquor and lignosulfonates, two water-soluble lignin forms from the pulp and paper industry \citep{Belkheiri_2018a,Rana_2019}. Lignin constitutes the remaining fraction after isolating the cellulose and hemicellulose in the pulping process and is an attractive feedstock from a biorefinery perspective. There are few processing solutions for this industrial by-product, and there is ample supply worldwide (up to 100 million tonnes/year in 2015 \citep{Bajwa_2019}). The pulping process efficiency can also improve from lignin down-processing and extraction technologies, contributing to their portfolio diversification and revenue generation \citep{Dessbesell_2020}.

While lignin that originates from pulping processes is water-soluble, it presents a shear-thinning behaviour \citep{Vainio_2008,Costa_2011}, and so aqueous lignin mixtures are a type of non-Newtonian fluid.  These fluids' rheology and flow dynamics present a challenge, as  data  is difficult to obtain. The high temperature and pressure of the hydrothermal medium constitute a challenge to carry out any experimental studies, and there is thus merit in a computational approach.
The current state of the art models for aqueous biomass mixtures consider low solid concentrations, so thermophysical properties can be assumed not to differ significantly from pure water properties \citep{Cantero_2013, Tran_2017a, Ranganathan_2018}. This approach severely limits their application to real-world situations, where it is estimated that a concentration above $36.6\%_{wt.}$ is necessary to ensure the economic feasibility of the HTL process \citep{Knorr_2013}. Therefore, simulations assuming a dilute feed might incur significant errors.

Regression models can derive a thermophysical description of biomass and water mixtures. \Citet{Schneider_2016} developed density, heat capacity and viscosity equations for algae and water mixtures based on this approach. The resulting model is a function of temperature and solids concentration (up to 335 K and 20 wt.\%, respectively). However, HTL operates at significantly higher pressures, temperatures and solid concentrations. Any extrapolation will lead to thermophysical property values with high uncertainty.

Knowing and predicting non-Newtonian fluid properties and flow patterns at HTL conditions allows the study of suitable reactor configurations and comprehensive process optimisation, all pivotal to achieving industrial-scale operation. For example, ensuring adequate mixing will positively impact both heating rate and residence times, thus minimising unwanted secondary reactions \citep{Faeth_2013, Bach_2014, Tran_2016, Hietala_2016, Qian_2017}. Also, computing residence time distribution (RTD) curves allow assessing reactor geometry design performance using a well-known and established metric within the chemical engineering field.
Both these goals require an adequate thermophysical characterisation of the fluid.

This study develops a thermophysical model applicable for shear thinning, water-soluble lignin mixtures, where a power law describes their viscosity. The model can predict the behaviour of non-Newtonian fluid flows in sub and supercritical water conditions and at solid loadings relevant for industrial applications. The flow simulations considered a continuous stirred tank reactor (CSTR) and a plug flow reactor (PFR) in series to  compute RTD curves, mixing time ratios, and frozen adiabatic mean temperatures, all response variables relevant for reactor design. The  uncertainty quantification procedure assesses the sensitivity of the results to any thermophysical model parameters and defines a confidence interval for the selected response variables.

\section{Materials and Methods}\label{sec:methodology}
This section explains the methodology for obtaining the thermophysical model, the uncertainty quantification procedure and its implementation.
\subsection{Thermophysical properties}\label{ssec:props}
The comprehensive assessment of the reactor's performance requires a  thermophysical model of biomass mixtures in water, expressed as a function of temperature, pressure and solid concentration.
However, due to difficulties in obtaining experimental data, the model cannot incorporate the contributions of all state variables.  Thus, pressure effects are only considered for the pure water terms, as these are calculated by equations of state, following the well-established IAPWS95 formulation \citep{Wagner_2002}.

The heterogeneous nature of lignin as a feedstock calls for a different approach to model parametrisation, especially regarding biomass-related equation terms. The novelty of this work lies in the use of a set of \textit{uncertain parameters}, each assuming a probability distribution function representing the natural variability of lignin properties due to being sourced from a hardwood or softwood, different locations, how and when the wood was harvested, and the processing type and conditions to obtain the concentrated lignin feedstock. As is customary in most modelling approaches, the remaining model \textit{fixed parameters} take a single value. The uncertain parameters tie into the uncertainty quantification (UQ) methodology described in \cref{ssec:uq}.

\subsubsection{Mixing laws}\label{sssec:mix-laws}

The lignin mixture's properties are described by mixing laws, usually a weighted average between water and solid properties \citep{Chhabra_2008}. In the case of mixture density, it becomes

\begin{equation}
	\rho_{\text{mix}} = \rho_w (1-\varphi)+ \mathcal{K}_{\rho}\varphi \ \text{,} \label{eq:mix_rho}
\end{equation}
where $\mathcal{K}_{\rho}$ is the lignin density, directly considered as an uncertain parameter, as it is not a function of any state variable and is the most straightforward approach that allows the use of the UQ methodology described in \cref{ssec:uq}.

The mixture heat capacity is defined as
\begin{align}
	C_{P,\text{mix}} &= C_{P,w} (1-\varphi) + \mathcal{K}_{f_{H}} \Phi_{C_P}\varphi \ \text{,} \label{eq:mix_cp}\\
\intertext{where}
	\Phi_{C_P} &=  C_{h,0} + C_{h,1} T \ \text{.} \label{eq:phi_cp}
\end{align}
$\mathcal{K}_{f_H}$ is a multiplicative factor that controls the relative influence of lignin on the overall mixture heat capacity.
\Cref{eq:phi_cp} is a polynomial that describes the temperature relation of solid lignin's heat capacity, and it is defined by the constants $C_{h,0}$ and $C_{h,1}$. The same equation, with the same values for the constants, was used in a previous study to predict the heat capacity of lignin-based compounds \citep{Gorensek_2019}.
$\varphi$ corresponds to the mass fraction of solids and subscripts ``$\text{w}$'' and ``$\text{mix}$'' identify water and mixture properties, respectively.

\subsubsection{Constitutive laws}\label{sssec:const-laws}

A power-law equation can describe the shear-thinning behaviour of biomass and water mixtures  \citep{Stickel_2009}. Additionally, it is one of the simplest viscosity constitutive laws, applicable to a wide range of fluids \citep{Chhabra_2008}. The power-law viscosity constitutive law is commonly presented in the form

\begin{equation}
	\eta=K\dot{\gamma}^{n-1} \ \text{,} \label{eq:power_law}
\end{equation}
where $K$ represents the consistency index, and $n$ is the power-law index. These two parameters are obtained by linearising \cref{eq:power_law} and performing a regression using rheology measurements - shear stress (or viscosity) as a function of shear rate. However, the obtained regressed parameters are specific to the conditions they were measured, while in this study, the goal is to derive expressions applicable to a wide range of temperature, pressure and concentrations. To this effect, models for the power-law coefficients, $K$ and $n$ need to be developed.

Considering the consistency index as the shear independent component of viscosity \citep{Mueller_2011}, then, $K_\text{mix}$ can be expressed as the weighted average between water and lignin shear-independent viscosity:

\begin{equation}
	K_\text{mix} = \eta_w (1 - \varphi) + \eta_0\varphi \ \text{,} \label{eq:mix_K}
\end{equation}
where $\eta_w$ and $\eta_0$ are the water and zero-shear viscosity \citep{Vainio_2008}, respectively.

The semi-empirical Vogel, Tamman and Fulcher equation was chosen as the basis to model $\eta_0$. A previous study used this expression to determine the shear-independent contribution term to the viscosity of a lignin mixture \citep{Costa_2011}, combined  with a modified Quemada equation. This work extends its application to power-law fluids. The original equation is

\begin{equation}
	\eta_0=A \exp \left(\frac{B}{T-T_0}\right) \ \text{,} \label{eq:VTF}
\end{equation}
where $A$ and $B$ are analogous to the pre-exponential factor and activation energy in an Arrhenius equation, respectively, and $T_0$ to the glass transition temperature.
Replacing the parameters $B$ and $T_0$ by concentration-dependent expressions~\citep{Costa_2011} and adding a multiplicative factor $\mathcal{K}_{f_B}$ to the exponential part of the equation yields

\begin{equation}
	\eta_0 = A\exp\left(\mathcal{K}_{f_B}\frac{B_1+B_2\varphi}{T-C\varphi}\right) \ \text{.} \label{eq:eta0}
\end{equation}
The parameters $A$, $B_1$, $B_2$ and $C$ in \cref{eq:eta0} remain fixed, while $\mathcal{K}_{f_B}$ is considered the uncertain parameter. Similarly to $\mathcal{K}_{f_H}$ in \cref{eq:phi_cp}, $\mathcal{K}_{f_B}$ will introduce random variability to the $\eta_0$ final value, making it either more or less sensitive to the effects of temperature and solids concentration.

 The mixture power-law index, $n_\text{mix}$, is a function of biomass concentration, modelled based on an approach using percolation theory~\citep{Campbell_2018a, Campbell_2018b}. This theory relates the motion of the solid particles within a fluid with its geometrical characteristics (e.g. particle shape and aspect ratio). Macroscopic transport properties such as mixture viscosity can be modelled based on fundamental physical principles and therefore extend the prediction range of such equations, especially in comparison to purely stochastic correlations.

\begin{figure}
	\centering
	\includegraphics[width=0.55\textwidth]{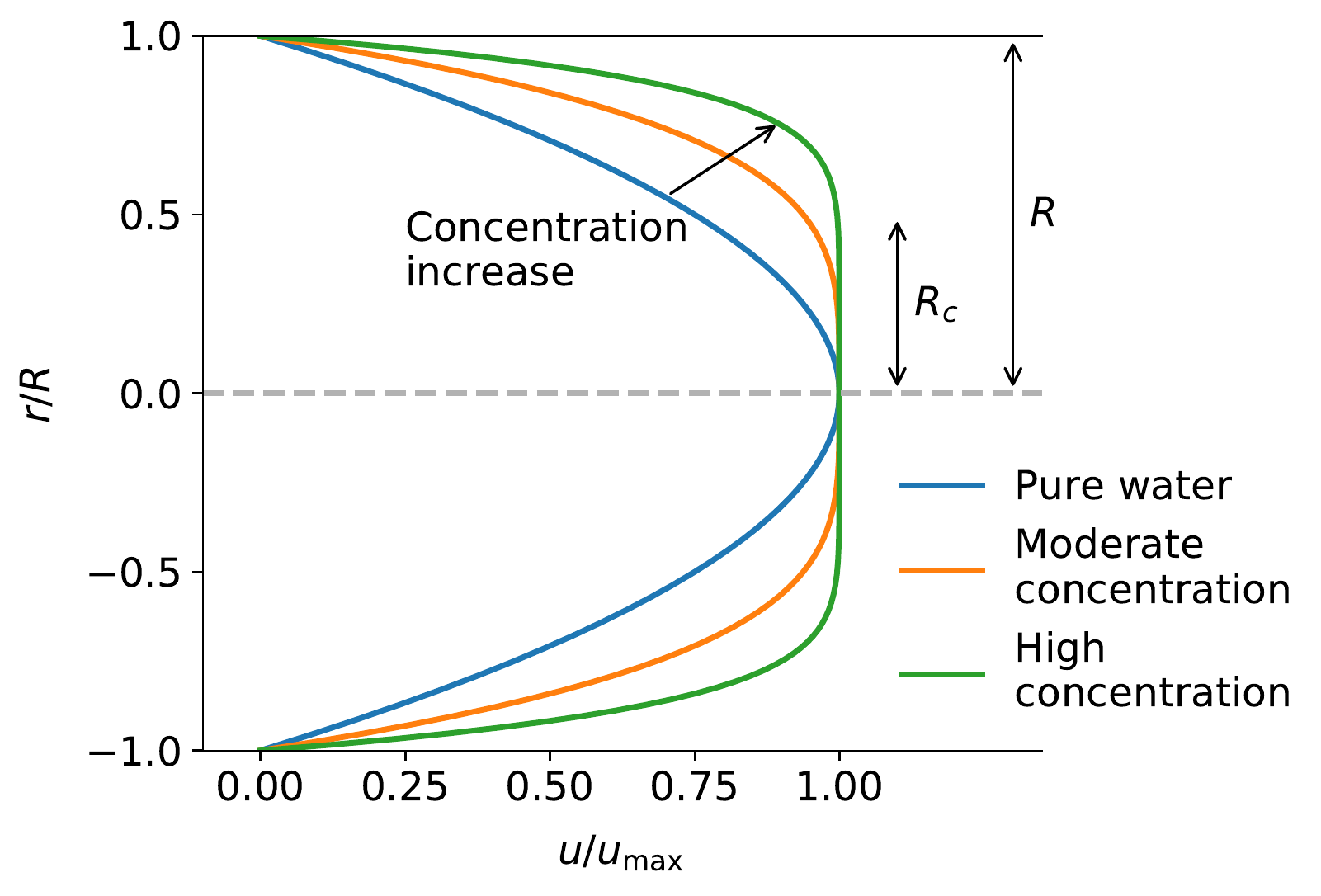}
	\caption{Schematic of how the velocity profile changes with increasing solid concentration. The velocity profile changes from parabolic to partial plug flow as solid concentration increases. In the centre, the fluid shows plug flow behaviour, while near the walls, the velocity steadily decreases until reaching zero (at the wall). The quantities used to define the plug size are presented as $\mathrm{R}_c$ and $\mathrm{R}$, the radius of the plug and pipe, respectively \citep{Campbell_2018a}.}
	\label{fig:plug_size}
\end{figure}

\Citet{Campbell_2018b} define the power-law index as

\begin{equation}
	n_\text{mix}=1-\frac{4}{3} \xi+\frac{1}{3} \xi^{4} \ \text{,} \label{eq:n}
\end{equation}
where $\xi$ corresponds to a quantity defined as the plug size. Considering the velocity parabolic profile of a Poiseuille flow, as the fluid starts to exhibit non-Newtonian behaviour, the velocity at the  pipe core will begin to flow as a plug. In contrast, near the wall, velocity will still tend to zero as in a regular Poiseuille flow (see \cref{fig:plug_size}). The expression to determine the plug size based on particle characteristics \citep{Campbell_2018a} is

\begin{equation}
	\xi=\frac{R_{\mathrm{c}}}{R}=\left[\frac{\phi-\phi_c}{\phi_m-\phi_c}\right]^{\mathcal{K}_{\beta_{2}}} \ \text{,} \label{eq:plug_size}
\end{equation}
where $\mathrm{R}_c$ and $\mathrm{R}$ correspond to the plug and pipe radius, respectively, $\phi$ is the volumetric fraction, $\phi_m$ is the maximum packing fraction and $\phi_c$ the critical volume fraction. The geometrical arguments used to derive the equations imply that volumetric fractions must be used when computing $n$.
The exponent $\mathcal{K}_{\beta_{2}}$ controls the growth of the plug size with solid concentration and is dependent on factors such as particle size, particle size distribution, presence of non-hydrodynamic forces such as Brownian motion and the fractal dimension of the clusters \citep{Campbell_2018a}.

\Citet{Mueller_2011} defined $\phi_m$ as a function of the particle aspect ratio:

\begin{equation}
	\phi_{\mathrm{m}}=\phi_{\mathrm{m}_1} \exp \left[-\frac{\left(\log_{10} A_{\mathrm{p}}\right)^{2}}{2 b^{2}}\right] \ \text{,} \label{eq:max_pack}
\end{equation}
where $\phi_{\mathrm{m}_1}$ is the maximum packing fraction in the case of perfect spheres, and $A_\mathrm{p}$ is the particle aspect ratio. \Cref{eq:max_pack} takes the form of a log-normal function with unity as mean and $b$ as the standard deviation that modifies the value of $\phi_{\mathrm{m}_1}$, as it deviates from the ideal case of perfect spheres. The value of $\phi_{\mathrm{m}_1}$ is 0.64, confirmed both experimentally and computationally. The value of $b$ used in this study, $1.171$, is given by \citet{Klein_2018}, which followed the original work of \citet{Mueller_2011} and performed a new curve fitting with a larger dataset.

\Citet{Vovchenko_2011} derived the following expression for $\phi_c$:

\begin{align}
	\phi_c &= 1-\exp \left(-\left\langle V_{ex}\right\rangle \frac{V}{\left\langle v_{e x}\right\rangle}\right) \label{eq:crit_vol} \ \text{,}\\
\intertext{where}
	{\left[\frac{V}{\left\langle v_{e x}\right\rangle}\right]}^{-1} &= 2+\frac{3}{2}\left(1+\frac{\sin ^{-1} \varepsilon}{\varepsilon \sqrt{1-\varepsilon^{2}}}\right)\left(1+\frac{1-\varepsilon^{2}}{2 \varepsilon} \ln \frac{1+\varepsilon}{1-\varepsilon}\right) \label{eq:vex_ratio}\\
\intertext{and}
	\varepsilon &= \sqrt{1-A_{\mathrm{p}}^{-2}} \ \text{,} \label{eq:part_ecc}
\end{align}
where $\left\langle v_{e x}\right\rangle$ and $\left\langle V_{ex}\right\rangle$ are the average and critical total average excluded volumes, respectively, $V$ is the single  particle volume and $\varepsilon$ is the particle eccentricity.
$\left\langle V_{ex}\right\rangle$ is also a function of aspect ratio and is determined by interpolation using the data from \citep{Vovchenko_2011}. By replacing \cref{eq:vex_ratio,eq:part_ecc} in \cref{eq:crit_vol}, $\phi_c$ can be expressed solely as a function of $A_{\mathrm{p}}$.

The particle aspect ratio, $A_{\mathrm{p}}$, is obtained by solving the following expression for $[\eta]$:

\begin{align} \label{eq:aspect_ratio}
	\left[\eta\right] &= -\frac{1}{5 A_{\mathrm{p}}}+\lambda\left(1+0.058 \frac{(A_{\mathrm{p}}-1)^{2}}{A_{\mathrm{p}}}-0.029(\ln A_{\mathrm{p}})^{2}\right)  +\frac{4 A_{\mathrm{p}}^{2}}{5 \ln \left(1+A_{\mathrm{p}}^{3}\right)} \ \text{,}\\
\intertext{where}
	\lambda &= \frac{27}{10} - \frac{4}{5 \ln{2}} \ \text{,} \label{eq:lambda}
\end{align}
and $[\eta]$ is the intrinsic viscosity \citep{Groot_2012}. $[\eta]$ can be related to the particle molecular weight by the Mark-Houwink equation \citep{Braaten_2003}:

\begin{equation}
	\left[\eta\right]=K_{MH} \ \mathcal{K}_{M_W}^{\alpha_{MH}} \ \text{,} \label{eq:MH_eq}
\end{equation}
where $\mathcal{K}_{M_w}$ is the lignin molecular weight and $K_{MH}$ and $\alpha_{MH}$ are parameters specific to each biomass mixture.

Looking at \cref{eq:n,eq:plug_size,eq:max_pack,eq:crit_vol,eq:vex_ratio,eq:part_ecc,eq:aspect_ratio,eq:lambda,eq:MH_eq}, the final value of $n_\text{mix}$ is dependent on the solids volume fraction, $\phi$, the exponent, $\mathcal{K}_{\beta_{2}}$, and weight-averaged molecular weight, $\mathcal{K}_{M_W}$, as both $\phi_c$ and $\phi_m$ can be expressed as a function of the latter. Determining a single value for $\mathcal{K}_{\beta_{2}}$ and $\mathcal{K}_{M_W}$ representative of most lignin mixtures poses a challenge. The former represents an aggregate of factors related to the solid particles and their interactions with the surrounding fluid \citep{Campbell_2018a} which is difficult to obtain measures of. The latter can vary several orders of magnitude for the same type of lignin \citep{Braaten_2003}. Therefore, $\mathcal{K}_{\beta_{2}}$ and $\mathcal{K}_{M_W}$ were considered the uncertain parameters when computing $n_\text{mix}$.

Combining \cref{eq:mix_K,eq:eta0} to compute $K_\text{mix}$ and \cref{eq:n,eq:plug_size,eq:max_pack,eq:crit_vol,eq:vex_ratio,eq:part_ecc,eq:aspect_ratio,eq:lambda,eq:MH_eq} to compute $n_\text{mix}$, the final viscosity equation is
\begin{equation}
	\eta_{\text{mix}} = \left[\eta_w (1 - \varphi) + \eta_0\varphi \right] \dot{\gamma}^{n-1} \ \text{.} \label{eq:visc_law}
\end{equation}
The Stokes-Einstein equation \citep{Edward_1970} is used to define the molecular diffusivity of the mixture:
\begin{align}
	\mathcal{D}_\text{mix} &=  \frac{k_{B}T}{6\pi r_p\eta_{\text{mix}}} \ \text{,} \label{eq:mol_diff}\\
\intertext{where}
	r_p &= m \ \mathcal{K}_{M_W}^p\label{eq:phi_D}
\end{align}
is the particle radius, $k_{B}$ the Boltzmann constant, $\eta_{\text{mix}}$ the fluid viscosity and $m$ and $p$ are fitting parameters specific to each type of lignin. \Cref{eq:phi_D} is a function of molecular weight, eliminating the need to define additional uncertain parameters for this thermophysical property.

All the relevant thermophysical properties, $\rho_\text{mix}$, $C_{P,\text{mix}}$, $\eta_{\text{mix}}$ and $\mathcal{D}_\text{mix}$ are now described by equations that are sensitive to key operation conditions in a chemical reactor: temperature, pressure, solid concentration and flow rate. The latter is convertible to shear rate, and so it will only be relevant to viscosity and indirectly to diffusivity calculations.

The uncertain parameter distributions and fixed parameter values considered in this study are shown in \cref{tab:up_dists,tab:fp_values}.

\begin{table}
	\centering
	\caption{Uncertain parameter (UP) final distributions used. Depending on the distribution, $\mu$, $\sigma$, $Z$, $s$, $a$ and $b$ are the distributions mean, standard deviation, shift, scale, lower and upper limits, respectively.}
	\begin{tabular}{lllrrrr}
		\toprule
		UP & UP name & Distribution & $\mu$ & $\sigma$ & $Z$ or $a$ & $s$ or $b$ \\
		\midrule
		$\mathcal{K}_{\rho}$ & Density & Normal & 1400  & 50    & - & - \\
		$\mathcal{K}_{f_H}$ & Heat Capacity Factor & Normal & 1     & 0.2   & - & - \\
		$\mathcal{K}_{f_B}$ & Activation Energy Factor & Uniform & - & - & 0.9   & 1.1 \\
		$\mathcal{K}_{\beta_2}$ & Plug Size Exponent & Uniform & - & - & 2     & 7 \\
		$\mathcal{K}_{M_W}$ & LS Molecular Weight & Lognormal & 10.1  & 1.0   & 59.0  & 0.56 \\
		$\mathcal{K}_{M_W}$ & BL Molecular Weight & Lognormal & 11.1  & 1.2   & 0.0   & 0.51 \\
		\bottomrule
	\end{tabular}
	\label{tab:up_dists}
\end{table}

\begin{table}
	\centering
	\caption{Fixed parameter final values used, indicating the respective property and equation where they appear. BL and LS are black liquor and lignosulfonates mixtures, respectively.}
	\begin{tabular}{llrrr}
		\toprule
		\multirow{2}[2]{*}{Property} & \multicolumn{1}{c}{\multirow{2}[2]{*}{Parameter}} & \multicolumn{2}{c}{Value} & \multirow{2}[2]{*}{Equation} \\
		&       & \multicolumn{1}{r}{LS} & \multicolumn{1}{r}{BL} &  \\
		\midrule
		Density & -     & - & - & - \\
		\multirow{2}[0]{*}{Heat Capacity} & $C_{h,0}$ & \multicolumn{2}{c}{0.064} & \multirow{2}[0]{*}{(\ref{eq:phi_cp})} \\
		& $C_{h,1}$ & \multicolumn{2}{c}{0.004} &  \\
		Viscosity &       &       &       &  \\
		\multicolumn{1}{r}{\multirow{4}[0]{*}{\textit{\makecell[r]{Power law\\index}}}} & $\beta_{2,\text{ref}}$ & 4.82 & 3.78 & (\ref{eq:plug_size}) \\ 
		& $K_{MH}$ & 0.120 & 0.079 & \multirow{2}[0]{*}{(\ref{eq:MH_eq})} \\
		& $\alpha_{MH}$ & 0.360 & 0.307 &  \\
		& $b$   & \multicolumn{2}{c}{1.171} & (\ref{eq:max_pack}) \\
		\multicolumn{1}{r}{\multirow{4}[0]{*}{\textit{{\makecell[r]{Zero-shear\\viscosity}}}}} & $A\times10^4$   & 8.156 & 5.8 & \multirow{4}[0]{*}{(\ref{eq:eta0})} \\
		& $B_1$ & 500.0 & 344.5 &  \\
		& $B_2$ & 500.0 & 461.2 &  \\
		& $C$   & 461.0 & 396.7 &  \\
		\multirow{2}[1]{*}{Diffusion} & $m$   & \multicolumn{2}{c}{1.067} & \multirow{2}[1]{*}{(\ref{eq:phi_D})} \\
		& $p$   & \multicolumn{2}{c}{0.281} &  \\
		\bottomrule
	\end{tabular}
	\label{tab:fp_values}
\end{table}

\subsection{Uncertainty quantification procedure} \label{ssec:uq}

\begin{figure}
	\centering
	\includegraphics[width=\textwidth]{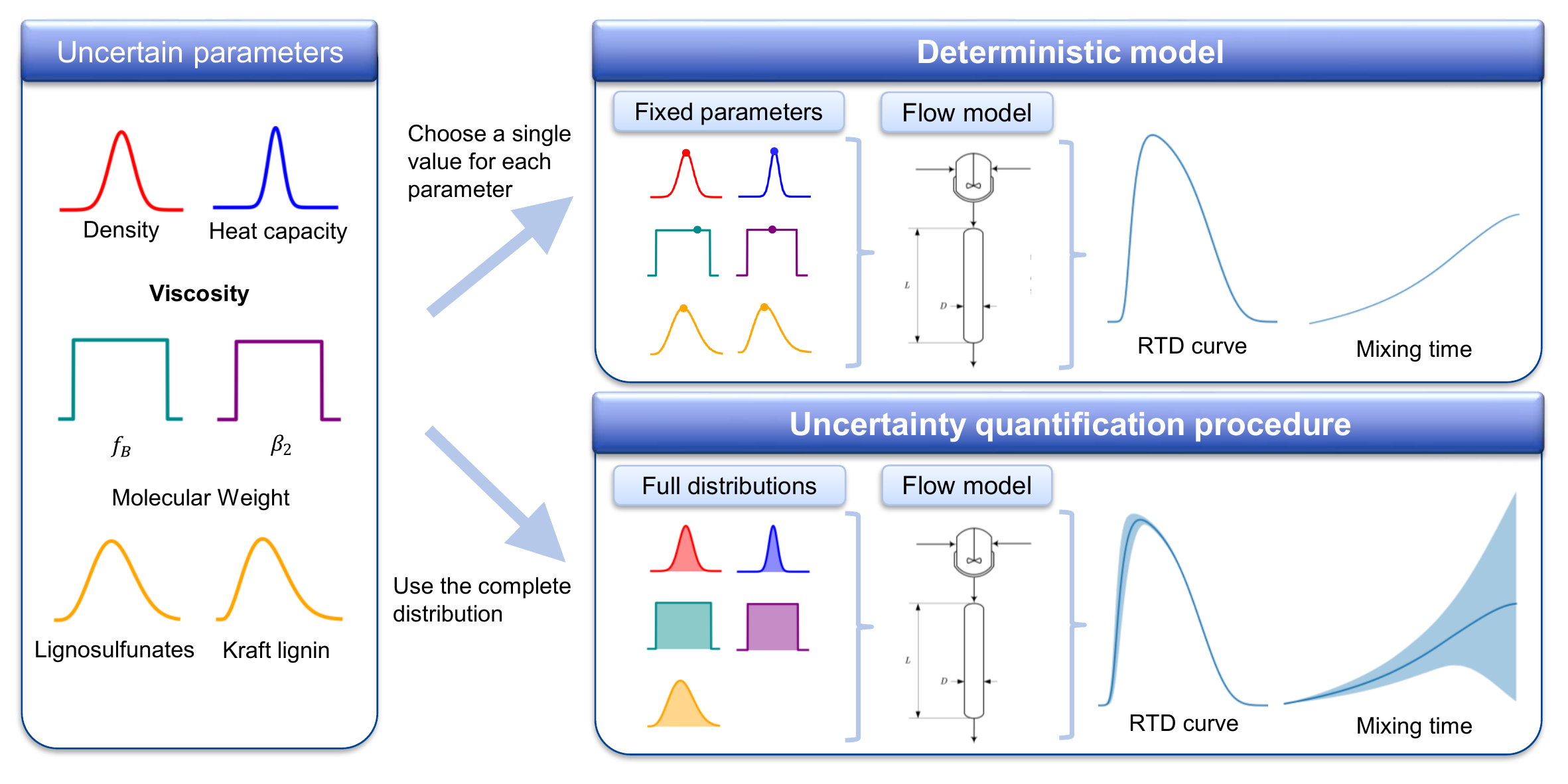}
	\caption{Schematic representation of the uncertainty quantification methodology and comparison to a deterministic model approach. The uncertain parameters assume a distribution fed into the flow model, allowing the determination of a confidence interval for the response variables. All curves are merely representative.}
	\label{fig:UQ_method}
\end{figure}

The UQ procedure quantitatively determines the influence of the thermophysical properties on the response variables or quantities of interest (QoI). Based on the probability distribution functions assumed by $\mathcal{K}_{()}$, this procedure will generate valid model inputs. The model can then take the single value parametric inputs and compute the QoIs. The UQ procedure uses polynomial chaos expansions \citep{Xiu_2005} where statistical metrics such as the mean, $\mathbb{E}$, variance, $\mathbb{V}$, percentiles, $P_x$, and the prediction interval, $I_x$, are determined for each QoI. If the  amount of uncertain parameters is below 20, this method is considerably faster than \mbox{quasi-Monte}~Carlo methods \citep{Xiu_2005, Crestaux_2009, Eck_2016}, which makes the methodology well suited for the current problem. \Cref{fig:UQ_method} shows a schematic representation of the UQ methodology.

The QoI, denoted here by the variable $Y$, can be the RTD curve, mixing time, mixing temperature, or any other output considered relevant to chemical reactor flow. The response variable is a function of several independent variables: total mass flow, hot and cold flow ratios, operation pressure, inlet temperatures, and solids concentration.
For $N$ independent variables such that $\boldsymbol{x}=\left[x_1, x_2, \ldots, x_N\right]$ and $d$ uncertain independent input parameters $\boldsymbol{\mathcal{K}}=\left[\mathcal{K}_1, \mathcal{K}_2, \ldots, \mathcal{K}_{d}\right]$, the output $Y$ is

\begin{equation}
	Y=U(\boldsymbol{x},\boldsymbol{\mathcal{K}}) \ \text{,}
\end{equation}
where $Y$ can have any value within the output space $\Omega_Y$ and has an unknown probability density function $\varrho_Y$. The goal of the UQ procedure is to compute $\varrho_Y$ and simultaneously describe the influence of the uncertain parameters on the QoI. The first-order Sobol indices can quantitatively describe the latter:

\begin{equation}
	S_{()}=\frac{\mathbb{V}\left[\mathbb{E}\left[Y \mid \mathcal{K}_{()}\right]\right]}{\mathbb{V}[Y]} \ \text{,} \label{eq:sobol}
\end{equation}
where $\mathbb{E}$ and $\mathbb{V}$ correspond to the mean and variance of $Y$. $\mathbb{E}\left[Y \mid \mathcal{K}_{()}\right]$ denotes the mean value of $Y$ for the cases where the uncertain parameter $\mathcal{K}_{()}$ is not varied. The variance of this value will inherently be lower than the total variance when all uncertain parameters are varied ($\mathbb{V}[Y]$). Therefore, the Sobol index is a measure of variance reduction when $\mathcal{K}_{()}$ remains unchanged.

\subsection{Implementation in Python} \label{ssec:python}

The thermophysical model is coded in Python 3.8, with pure water properties determined by the \texttt{CoolProp} module, which follows the IAPWS95 formulation \citep{Wagner_2002}. \texttt{chaospy} handles the generation of distribution values and Gaussian kernel density estimation (KDE) for the molecular weight curves. To determine the log-normal distribution curve parameters other than the mean \verb|scipy.stats.rv_continuous.fit| is used. The curve fittings were performed with \texttt{scipy}'s \verb|curve_fit| module - non-linear least squares regression.
The \texttt{uncertainpy} package \citep{Tennoee_2018} is used to perform all UQ computations. It provides a framework to perform UQ straightforwardly, providing the inputs for the flow model and reading its outputs, varying the values of the uncertain parameters following the assigned distribution.
\texttt{scipy}'s \texttt{solve$\_$ivp} was used as the ODE solver. The PFR section of the reactor was implemented in \texttt{fipy}, using a 1D unsteady convection-diffusion, finite volume formulation, parallelized. 

\section{Model and simulation setup}

\subsection{Thermophysical model fitting} \label{ssec:prop_fit}

To adequately represent both lignosulfonate (LS) and black liquor (BL) mixtures, a set of uncertain and fixed parameters is defined for each mixture. The fixed parameter values used can be directly taken from literature (\cref{eq:phi_cp}), set by curve fitting (\cref{eq:eta0}) or regression (\cref{eq:MH_eq,eq:phi_D}). Each uncertain parameter will follow a distribution determined in an \textit{ad hoc} procedure, using the limited available data. Once all fixed and uncertain parameters are defined, Monte~Carlo (MC) simulations assess the thermophysical property's prediction range. The thermophysical model is considered to be fitted when the predicted range captures the experimental data or, in case this is lacking, falls within the expected variability of each property.

Normal distributions were assumed for density and heat capacity, using the parameter's average value as the distribution mean and choosing a standard deviation that best captures the property variability. These properties should not differ between BL and LS mixtures, so they were assumed to be the same for both.

Viscosity modelling is particularly challenging, so the model fitting procedure entails a larger number of steps when compared to other thermophysical properties. The whole procedure is as follows:

\begin{enumerate}
	\item Molecular weight distributions, different for each mixture, are obtained by Gaussian kernel density estimation (KDE) of sampled data. The KDE mean and standard deviation are used in the simplified log-normal distributions to prevent non-physical results \citep{Vainio_2008,Fricke_1998}.
	\item Parameters $A$, $B_1$, $B_2$ and $C$ are obtained by curve fitting \cref{eq:eta0} to zero-shear viscosity data \citep{Vainio_2008,Alabi_2010}.
	\item The mean value of the $\mathcal{K}_{\beta_2}$ distribution is determined by curve fitting 
	\cref{eq:plug_size} to experimental power law index data \citep{Vainio_2008,Alabi_2010,Zaman_1995}.
	\item The parameters $K_{MH}$ and $\alpha_{MH}$ in \cref{eq:MH_eq} are taken from \citet{Braaten_2003} for LS and obtained by curve fitting with experimental power law index data in the case of BL \citep{Alabi_2010,Zaman_1995}.
	\item The lower and upper limits of the uncertain parameter uniform distributions $\mathcal{K}_{\beta_2}$ and $\mathcal{K}_{f_B}$ are $\{2,7\}$ and $\{0.9,1.1\}$, respectively. The mean value of $\mathcal{K}_{\beta_2}$ is between 4-5, so the distribution varies by two units above and below this interval. $\mathcal{K}_{f_B}$ limits are a $\pm 10\%$ deviation to the reference unity value.
	\item Monte-Carlo simulations assess the prediction range for both indices and viscosity itself.
	\item Both fixed parameters and distribution limits (percentage deviation values) are adjusted to best capture the experimental data points.
\end{enumerate}

\subsection{Distributions} \label{ssec:dists}

\begin{figure}
	\centering
	\begin{subfigure}{0.45\textwidth}
		\includegraphics[width=\textwidth]{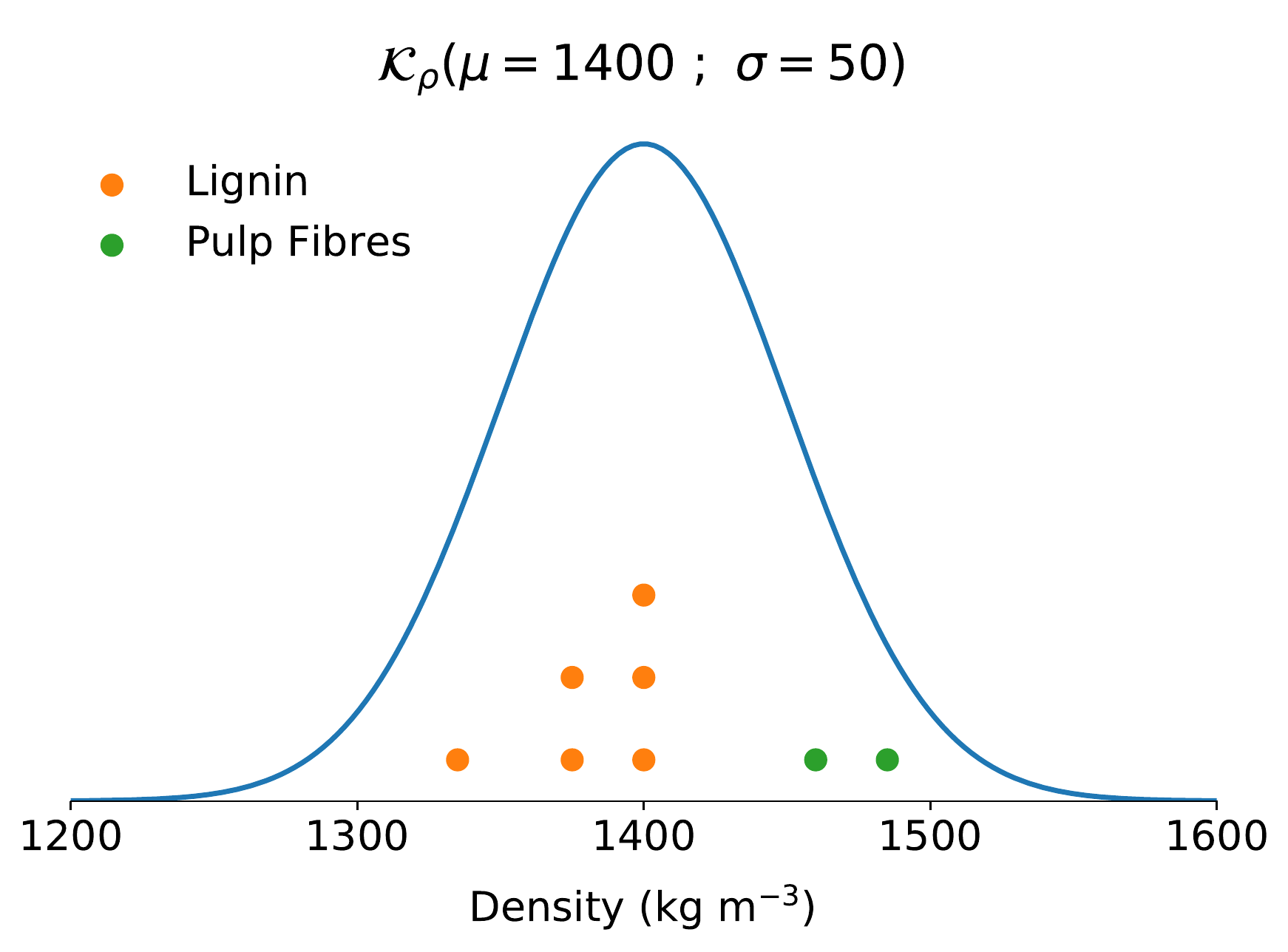}
		\caption{}
		\vspace{-0.9ex}
		\label{subfig:dist_rho}
	\end{subfigure}
	\begin{subfigure}{0.45\textwidth}
		\includegraphics[width=\textwidth]{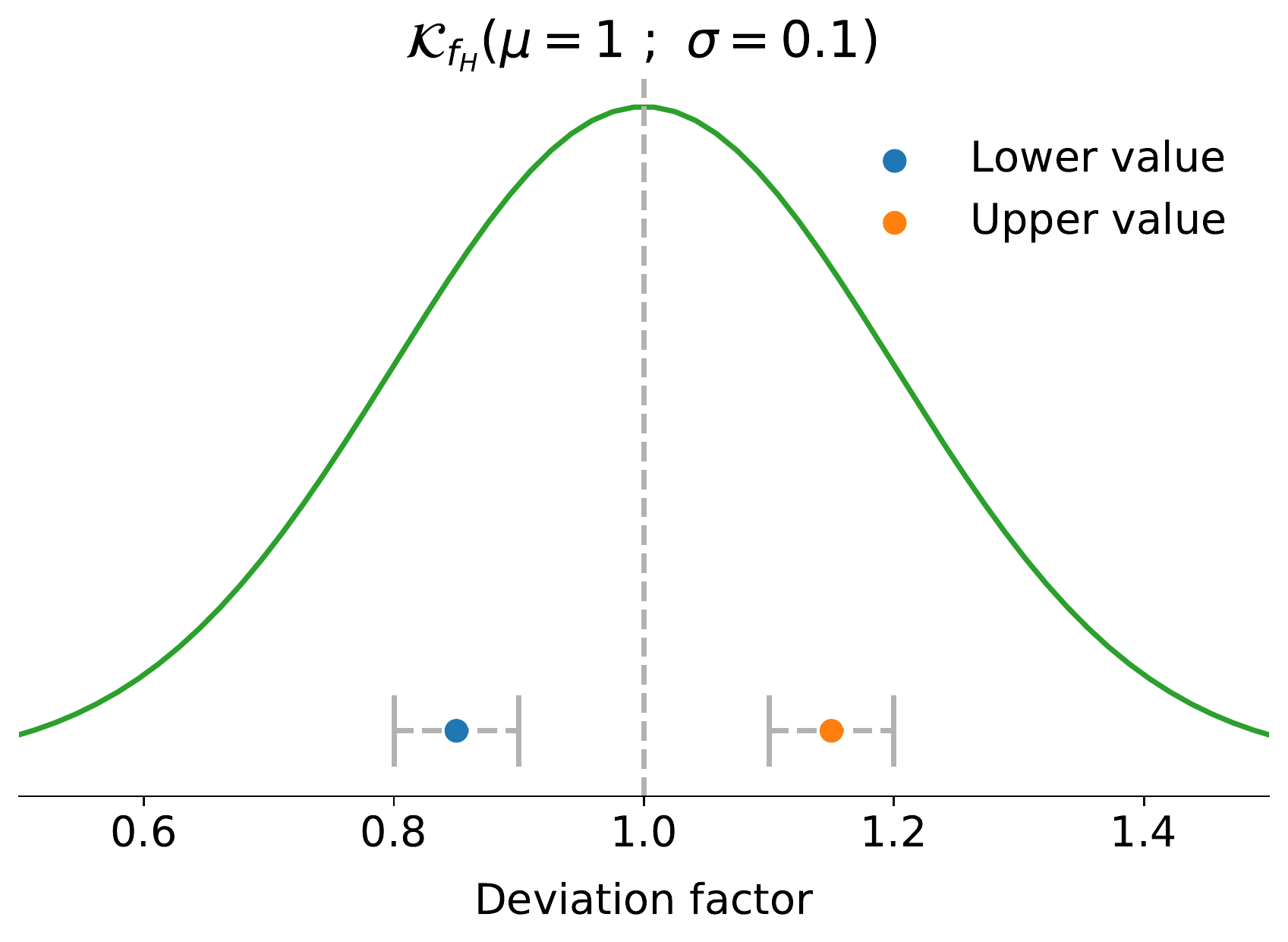}
		\caption{}
		\vspace{-1ex}
		\label{subfig:dist_cp}
	\end{subfigure}
	\caption{Density distribution function (a) and density values for lignin and wood pulp fibres \citep{Ehrnrooth_1984}. Deviation factor  distribution function (b) for heat capacity. The upper and lower values used to set the distributions standard deviations, along with error bars for each point, are shown \citep{Hatakeyama_2006}, as well as $\mathcal{K}_{f_H}$ mean (dashed line). The distribution function parameters (mean and standard deviation) are shown on top.}
	\label{fig:dist_rho-cp}
\end{figure}

\Cref{subfig:dist_rho} shows the probability distribution considered for density, along with experimental values for lignin and wood fibres. The density distribution average is $1400 \ \mathrm{kg/m^3}$, as per several sources in the literature \citep{Stamm_1929,Ehrnrooth_1984,Vainio_2008}. The standard deviation captures the experimental data of lignin and pulp fibres. The latter included to account for remaining wood impurities in the feedstock. Lignin density is lower than other wood components such as pulp fibres, hemicellulose or cellulose, so this distribution should account for several types of biomass impurities. The heat capacity deviation factor distribution function is shown in \cref{subfig:dist_cp}.  An  open literature  search found only three data points. The central point was considered the mean to generate the respective distribution. The remaining two values were considered an error bar. The final distribution captures all data points and their respective error bars.

\begin{figure}
	\centering
	\includegraphics[width=0.45\textwidth]{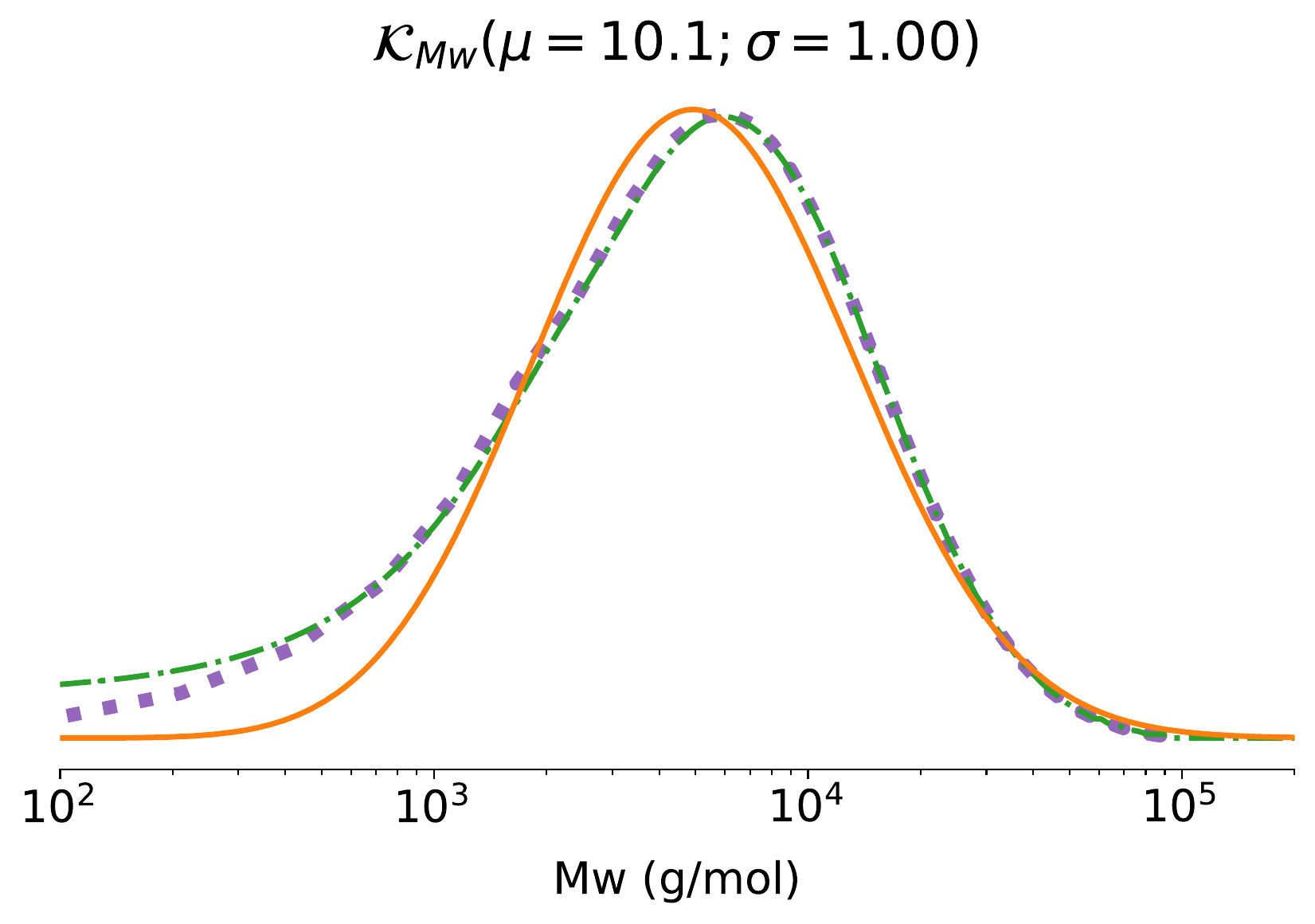}
	\hspace{2ex}
	\includegraphics[width=0.45\textwidth]{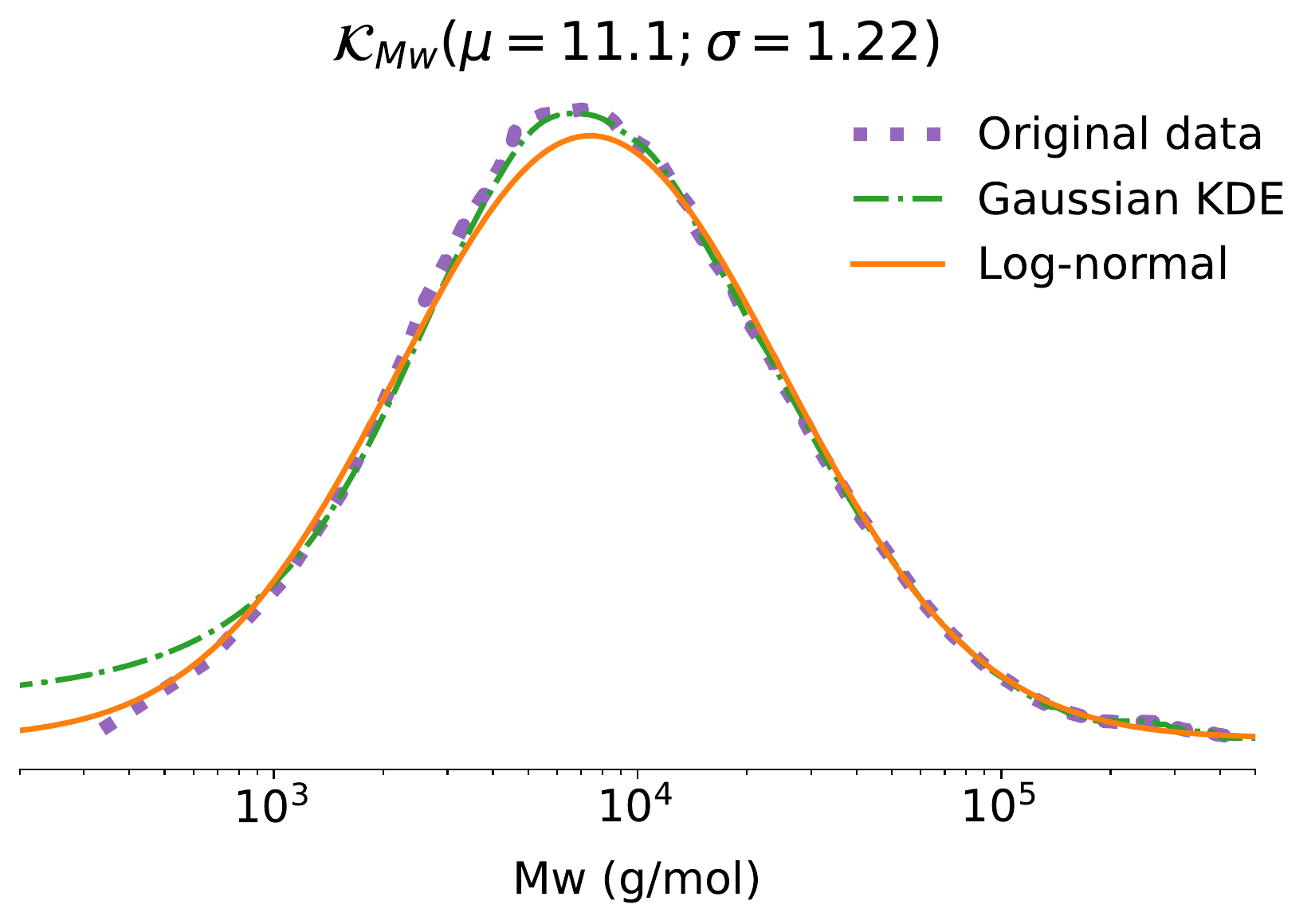}
	\caption{Lignosulfonates molecular weight (left) \citep{Vainio_2008}. Gaussian KDE bandwidth is 0.07. Black liquor molecular weight (right) \citep{Fricke_1998}. Gaussian KDE bandwidth is 0.1. The log-normal function parameters (mean and standard deviation) are shown on top.}
	\label{fig:dist_Mw}
\end{figure}

Molecular weight distributions for BL and LS are presented in \cref{fig:dist_Mw}. Two approaches were used, curve fitting assuming a log-normal function and Gaussian KDE. For  the KDE procedure to  be carried  out, the  original data was  sampled into histograms. Both approaches agree with the original data, except for values in the lower end of the molecular weight range. 
Accuracy loss from simplifying the distributions to a log-normal function is only significant at low molecular weights, for LS only.
Additionally, Gaussian KDE might interpret data irregularities as part of the distribution and can even take negative molecular weight values, causing errors to the UQ procedure. Therefore, the log-normal approximation  represents the molecular weight of both mixtures.

\subsection{Reactor flow model} \label{ssec:reactor}
\begin{figure}
	\centering
	\includegraphics[width=\textwidth]{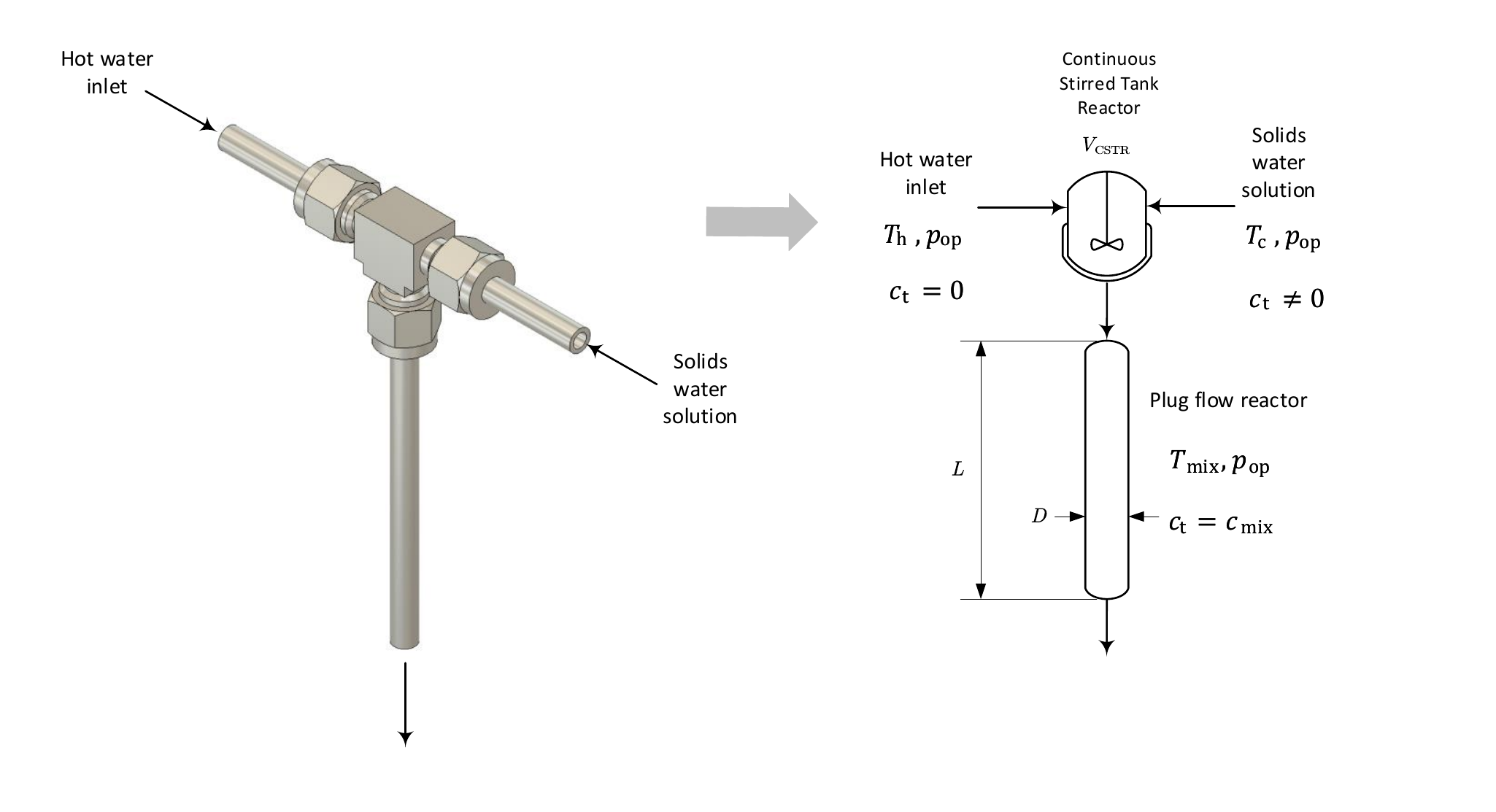}
	\caption{Diagram of the T-piece and pipe configuration used in this study (left). Model representation of the reactor (right). The inlets are mixed in the T-piece, pass through the plug flow section, and leave the reactor.}
	\label{fig:CSTR_PFR}
\end{figure}

To test the thermophysical model under some feasible scenario, \Cref{fig:CSTR_PFR} shows a schematic of the reactor simulated in this work and its model representation. The setup consists of a T-piece mixer followed by a pipe of length $L$. This geometry allows for high heating rates due to counter-current mixing between two streams at different temperatures \citep{Ikushima_2002,Blood_2004,Kawasaki_2010}. The hot water and aqueous solid solution inlets are assumed to mix in the T-piece section of the reactor instantaneously. The output of the mixer then goes through a plug flow reactor at isothermal conditions.

An idealised flow model consisting of a CSTR and a PFR in series is developed to represent this setup. A set of differential equations defines each section of the reactor. The CSTR section of the reactor is described by

\begin{align}
	\rho_{\text{mix}} V \frac{\partial c_{\text{tr}}}{\partial t} &= G_{c} c_{\text{tr,in}}-G_{\text{tot}} c_{\text{tr}} \label{eq:c_cstr} \\
\intertext{and}
	&\begin{aligned}
		\mathllap{\rho_{\text{mix}} {C_P}_{\text{mix}} V \frac{\partial T_{\text{mix}}}{\partial t}} &= \ G_{c} {C_P}_{c} T_{c}+G_{h} {C_P}_{h} T_{h} \\
		& \qquad - G_{\text{tot}} {C_P}_{\text{mix}} T_{\text{mix}}
	\end{aligned} \ \text{,} \label{eq:T_cstr}
\end{align}
where $c_{\text{tr}}$ corresponds to the concentration of a passive tracer, $c_{\text{tr,in}}$ to the tracer concentration in the CSTR inlet and $G$ are mass flow rates. The passive tracer concentration is followed to best mimic the experimental procedure to determine RTD curves, introducing a passive component in the system. The subscripts $\text{mix}$, $c$, $h$ and $\text{tr}$ denote mixture thermophysical properties and temperature, cold flow stream, hot flow stream and tracer concentration, respectively. Note that $\rho_{\text{mix}}$ and ${C_P}_{\text{mix}}$ depend on both temperature and solids concentration, which are constant throughout the reactor.

The concentration balance in the PFR part of the reactor configuration is
\begin{equation}
	\frac{\partial c_{\text{tr}}}{\partial t} + U \frac{\partial c_{\text{tr}}}{\partial x} = \mathcal{D}_{\text{eff}} \frac{\partial^2 c_{\text{tr}}}{\partial x^2} \ \text{,} \label{eq:c_pfr}
\end{equation}
where $U$ is velocity and $\mathcal{D}_{\text{eff}}$ is effective diffusion. $\mathcal{D}_{\text{eff}}$ is also a function of temperature and solids concentration. The passive tracer will not impact either the thermophysical properties values or state variables for both reactor sections.

\Cref{eq:c_cstr,eq:T_cstr,eq:c_pfr} are used for the RTD curve determination of the reactor. In the CSTR section, a system of ODEs (\cref{eq:c_cstr,eq:T_cstr}) determines the tracer concentration and mixture temperature. The PFR section is considered to be isothermal and so an unsteady one dimension convection-diffusion differential equation (\cref{eq:c_pfr}) is used to determine the final tracer concentration at the end of the reactor.

The effective diffusion, $\mathcal{D}_{\text{eff}}$, presented in \cref{eq:c_pfr}, can be defined as the sum of the mixture's molecular diffusivity ($\mathcal{D}_{\text{mix}}$), presented in \cref{eq:mol_diff} and a virtual coefficient of diffusion ($\varepsilon_\mathcal{D}$) related to flow advection, provided $\mathcal{D}_{\text{mix}}/\varepsilon_\mathcal{D} \ll 1$ \citep{Saffman_1960}:
\begin{equation}
	\mathcal{D}_{\text{eff}} = \mathcal{D}_{\text{mix}} + \varepsilon_\mathcal{D} \ \text{.} \label{eq:D_eff}
\end{equation}
The dispersion of a soluble material injected into a tube where a slow stream of a viscous fluid is flowing is defined by
\begin{equation}
	\varepsilon_\mathcal{D} = \frac{R^{2} U^{2}}{48 \mathcal{D}_{\text{mix}}} \ \text{.} \label{eq:eps_lam}
\end{equation}
In the case of turbulent pipe flow in smooth pipes, the virtual coefficient of diffusion takes the form of
\begin{align}
	\varepsilon_\mathcal{D} &= 7.14 R U \sqrt{\chi} \ \text{,} \label{eq:eps_turb}\\
\intertext{where}
	\chi^{-\frac{1}{2}} &=-0.40+4\log _{10} Re+2\log _{10} \chi \label{eq:turb_gamma}\\
\intertext{and}
	Re &= 2 R U / \nu \label{eq:reynolds}
\end{align}
are the fluid resistance correlation and the Reynolds number, respectively, with $\nu$ as the kinematic viscosity.

\Cref{eq:eps_lam,eq:eps_turb} are used depending on whether the flow regime is laminar or turbulent, respectively \citep{Taylor_1954}.
In both cases, $R$ and $U$ correspond to the pipe radius and flow velocity, respectively. \Cref{eq:eps_lam,eq:eps_turb,eq:turb_gamma} account for the flow conditions impact on the RTD curve, while the thermophysical effects are modelled by \cref{eq:mol_diff,eq:phi_D}.  

\subsection{Quantities of Interest} \label{ssec:qoi}

The \textit{mixing time} measures the micromixing degree between the two streams. A low micromixing time corresponds to better mixing between streams, which has a beneficial effect on the chemical reaction. Thus, how this quantity changes with the operating conditions reflects the overall reactor performance. \Citet{Baldyga_1989} define this quantity as
\begin{equation}
	\tau_{m}=12 \sqrt{\frac{\nu}{\epsilon}} \ \text{,} \label{eq:mix_time}
\end{equation}
where $\epsilon$ is the mean dissipation rate of turbulent kinetic energy and $\nu$ the kinematic viscosity.
Assuming $\epsilon$ is the same for a biomass containing stream and a pure water stream, the mixing time ratio can be defined as
\begin{equation}
	\frac{\tau_{m,s}}{\tau_{m,w}} = \sqrt{\frac{\nu_s}{\nu_w}} \ \text{.} \label{eq:mix_time_ratio}
\end{equation}
The mixing time ratio defined in \cref{eq:mix_time_ratio} compares the degree of micromixing attained by the simulated reactor configuration when considering the effect of biomass in the mixture against the pure water properties assumption.

The \textit{RTD curve} corresponds to the passive tracer concentration, $c_\text{tr}$, at the system outlet as a function of time, with both variables non-dimensionalized. To compute $c_\text{tr}$ over time, the PDE system formed by \cref{eq:c_cstr,eq:T_cstr,eq:c_pfr} must be solved, with $c_{\text{tr,in}}$ in \cref{eq:c_cstr} taking the value of a delta function. 

After solving the PDE system, the $c_\text{tr}(t)$ function must be non-dimensionalised, following the procedure listed in \citet{Levenspiel_2014} to allow for comparison with other RTD results.

The two streams are at the same pressure and mix adiabatically, instantaneously and at isobaric conditions. The mixing temperature, $T_\text{mix}$, is determined by an internal energy balance in \cref{eq:T_cstr} and follows the assumption of perfect mixing. This temperature can also be defined as the \textit{Frozen Adiabatic Mixing Temperature} (FAMT) \citep{Qiu_2015,Sierra-Pallares_2021}. 

\subsection{Test cases} \label{ssec:cases}

\begin{table}
	\centering
	\caption{Taguchi design of experiements table for the cases under study.}
	\begin{tabular}{lrrrrrrr}
		\toprule
		Case & $T_h$ & $F_{\text{tot}}$ & ${H/C}_{\text{ratio}}$ & \multicolumn{2}{c}{$Re$} \\
		& (K) & (g/s) &  & LS & BL \\
		\midrule
		F1    & 573.15 & 0.5   &  1:1  & 140   & 155 \\
		F2    & 623.15 & 0.5   &  4:3  & 310   & 254 \\
		F3    & 643.15 & 0.5   &  2:1  & 427   & 313 \\
		F4    & 573.15 & 2.5   &  4:3  & 1029  & 975 \\
		F5    & 623.15 & 2.5   &  2:1  & 1960  & 1473 \\
		F6    & 643.15 & 2.5   &  1:1  & 2270  & 1622 \\
		F7    & 573.15 & 12.5  &  2:1  & 6832  & 5799 \\
		F8    & 623.15 & 12.5  &  1:1  & 8740  & 6837 \\
		F9    & 643.15 & 12.5  &  4:3  & 12654 & 8675 \\
		\bottomrule
	\end{tabular}
	\label{tab:test_cases}
\end{table}

The hot water temperature and inlet flow rates were set based on the cases defined in \citet{Sierra-Pallares_2016a}, and the case numbers are sorted in ascending order of Reynolds number. The differences in this study are the addition of solids in the cold stream and an increase in total flow rate, resulting in higher Reynolds numbers. The reason for this change is to allow the effect of turbulence on effective diffusion to be studied by employing two different expressions to compute its value. Cases F1 to F3 are fully laminar, F4 to F6 are in a transitional regime, and the remaining cases are turbulent. \Cref{eq:eps_lam} and \cref{eq:eps_turb} are used to compute the coefficient of diffusion for cases F1 to F6 and F7 to F9, respectively. All test cases are listed in \cref{tab:test_cases}.

Only cases with inlet temperatures below the critical point of water were chosen for this study as salt solubility steadily drops when going from sub to supercritical water, at which point precipitation starts to occur, which can profoundly impact reaction yields. Most catalysts used in HTL reactions are homogeneous, soluble salts, meaning precipitation will negate their effect. Most HTL feedstocks ash content is high, BL and LS included, and while the role of these inorganics is not entirely understood, the indication is that they can have a beneficial role in HTL reactions. Additionally, the precipitated salts can cause corrosion and fouling issues to the equipment and may also catalyse thermal decomposition gaseous reactions, which is not the goal of HTL.

\section{Results and discussion} \label{sec:results}

Simulations with the thermophysical model are presented in the subsequent sections to illustrate the methodology's effectiveness. \Cref{ssec:pred_ranges} presents the model prediction ranges. \Cref{ssec:temp} presents the temperature results. \Cref{ssec:mixing} discusses the mixing time ratio, and \cref{ssec:RTD} the residence time distribution.

\subsection{Thermophysical model prediction ranges} \label{ssec:pred_ranges}

The thermophysical model fittings are first presented in prediction ranges or contours. Results for density, heat capacity, viscosity and its respective power-law parameters are shown, compared, when possible, with experimental data for the two-fluid mixtures studied. All prediction ranges correspond to 90\% confidence intervals, while the prediction contours for the power-law indices represent the raw simulation data. The uncertain parameter distributions and model fixed parameters are adjusted to capture the experimental results and provide physically consistent predictions. The reactor flow simulations are performed with the thermophysical model fitted for BL and LS, computing three QoIs: temperature, mixing time ratio, and RTDs. The operation conditions used correspond to the test cases presented in \cref{tab:test_cases}. The UQ procedure runs the simulations and computes, for all QoIs, a mean and a 90\% confidence interval based on polynomial chaos expansions and the respective Sobol indices.

\begin{figure}
	\centering
	\includegraphics[width=0.4\textwidth]{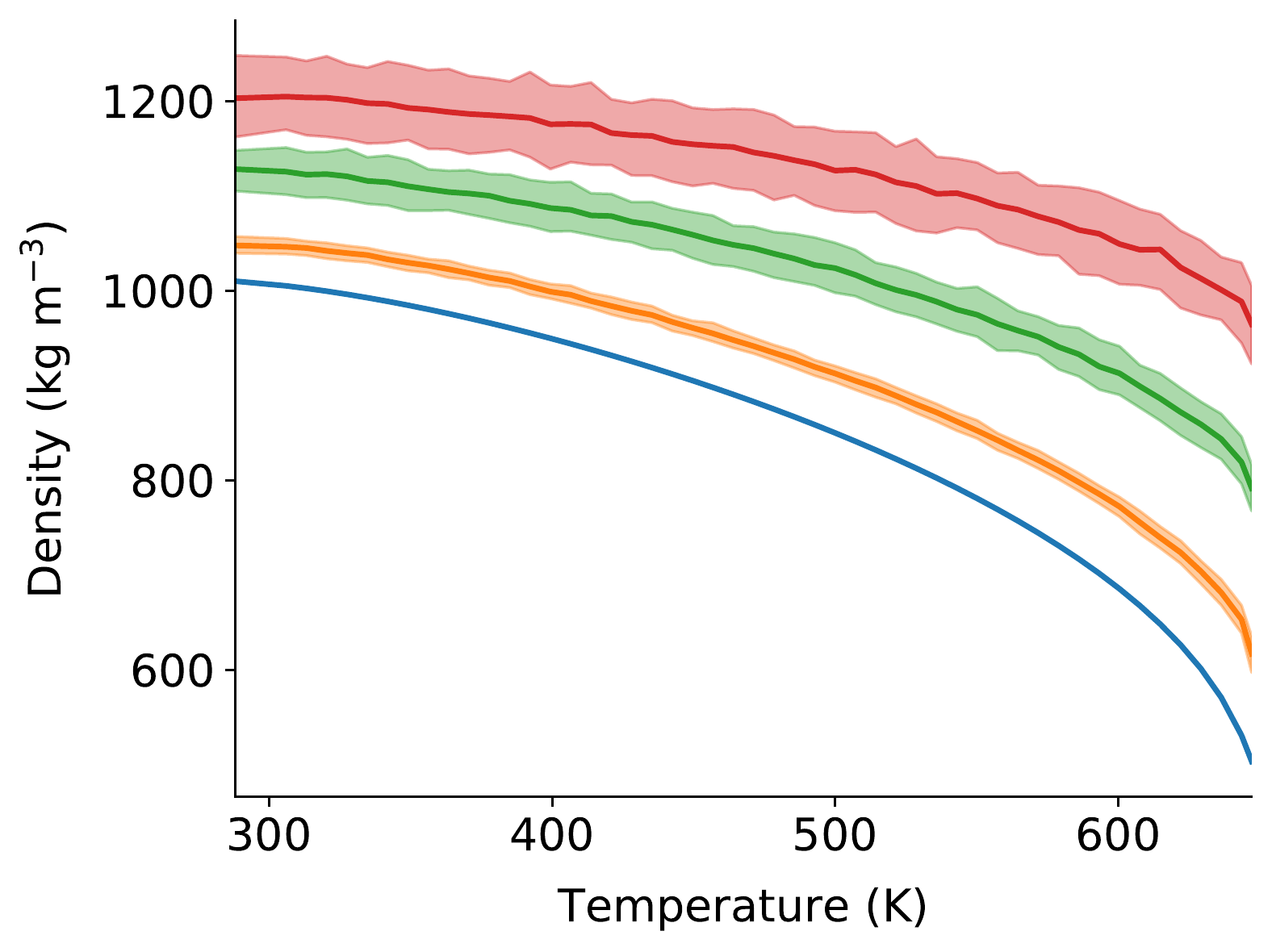}
	\hspace{2ex}
	\includegraphics[width=0.48\textwidth]{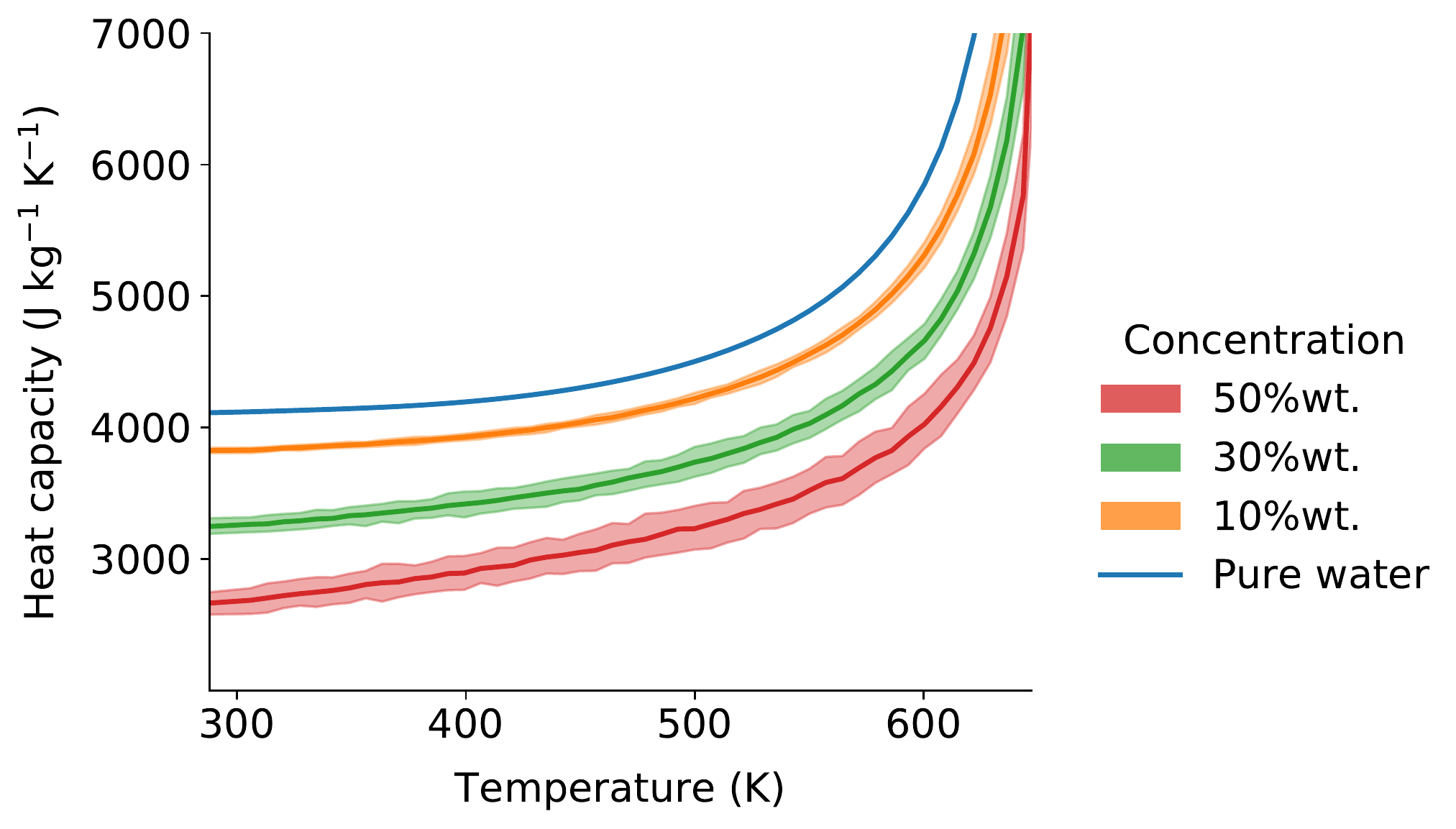}
	\caption{Mixture density (left) and heat capacity (right) as a function of temperature at different solid concentrations and respective $90\%$ confidence interval. Water properties (solid line) are shown as reference.}
	\label{fig:MC_props}
\end{figure}

\Cref{fig:MC_props} shows the results of the MC simulations for density and heat capacity for three different solid concentrations and as a function of temperature. Given that all these quantities are weighted averages of solid and water properties, the uncertainty increases with lignin concentration.
The heat capacity difference between pure water and lignin mixtures becomes exponentially smaller near the critical point, resulting  in an almost 5-fold heat capacity increase of water near this point, leading to a small relative contribution of the biomass polynomial to the final mixture weighted average heat capacity value.

\begin{figure}
	\centering
	\begin{subfigure}{.41\textwidth}
		\caption{}
		\vspace{-1ex}
		\includegraphics[width=\textwidth]{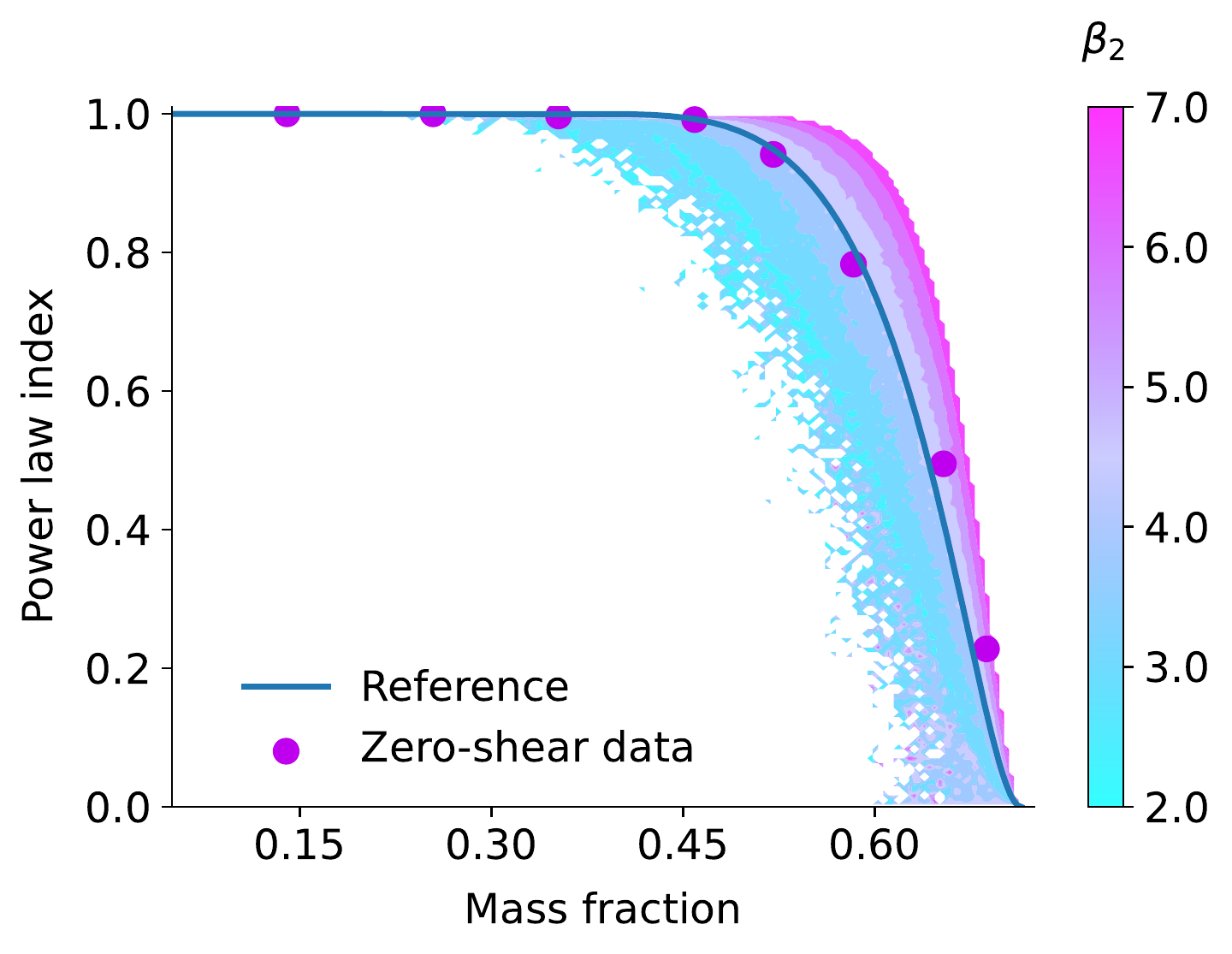}
	\end{subfigure}
	\begin{subfigure}{.45\textwidth}
		\caption{}
		\vspace{-1.5ex}
		\includegraphics[width=\textwidth]{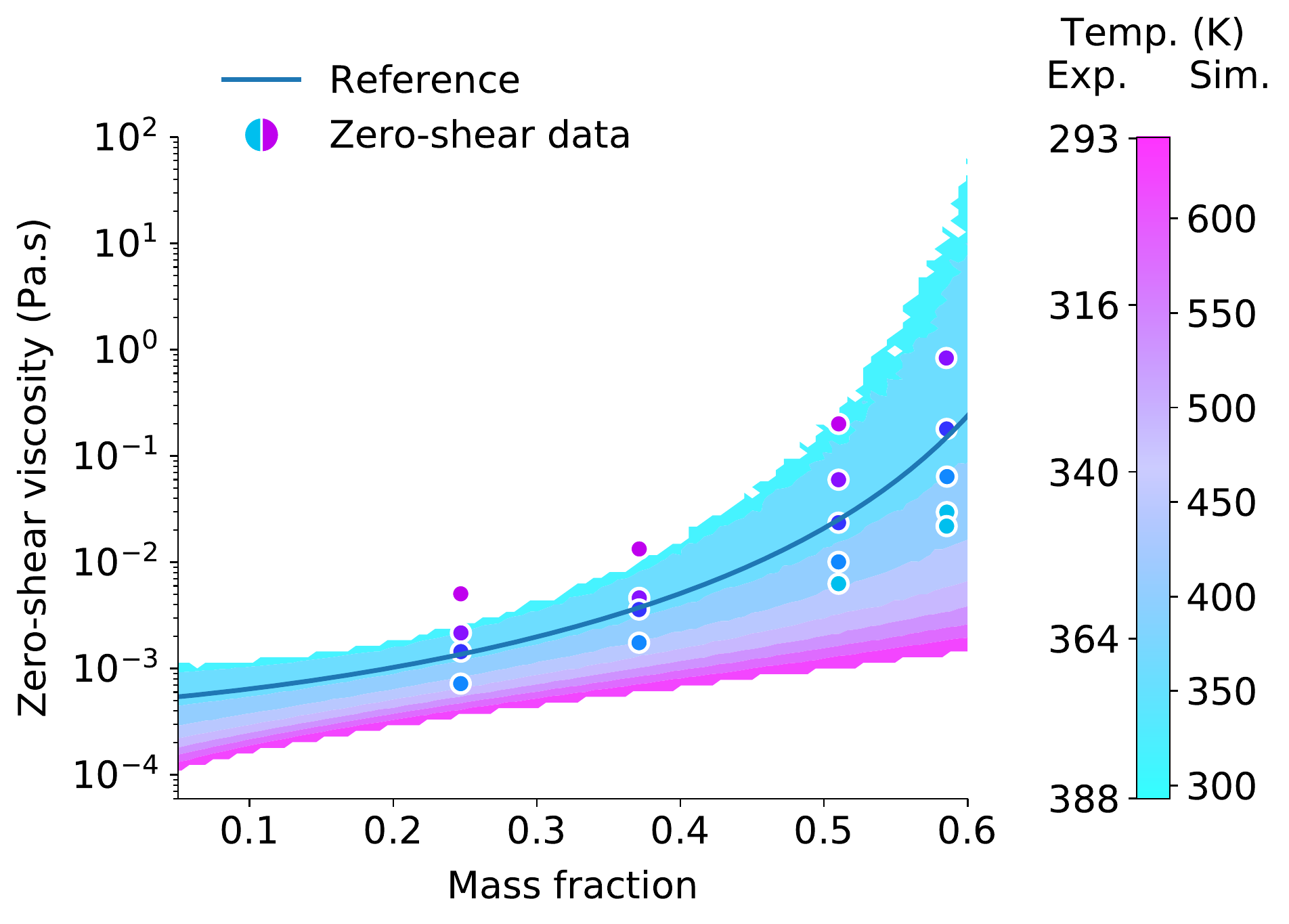}
	\end{subfigure}
	\caption{MC simulation results for the power-law index (a) and zero-shear viscosity (b) as a function of solids volume fraction (black liquor case). The contours show the power-law index and zero-shear viscosity dependence on $\beta_2$ and temperature, respectively. Experimental results and the model's prediction for the reference condition is also shown \citep{in Zaman_1995}.}
	\label{fig:MC_idx-BL}
\end{figure}

\begin{figure}
	\centering
	\begin{subfigure}{.45\textwidth}
		\caption{}
		\vspace{-1.5ex}
		\includegraphics[width=\textwidth]{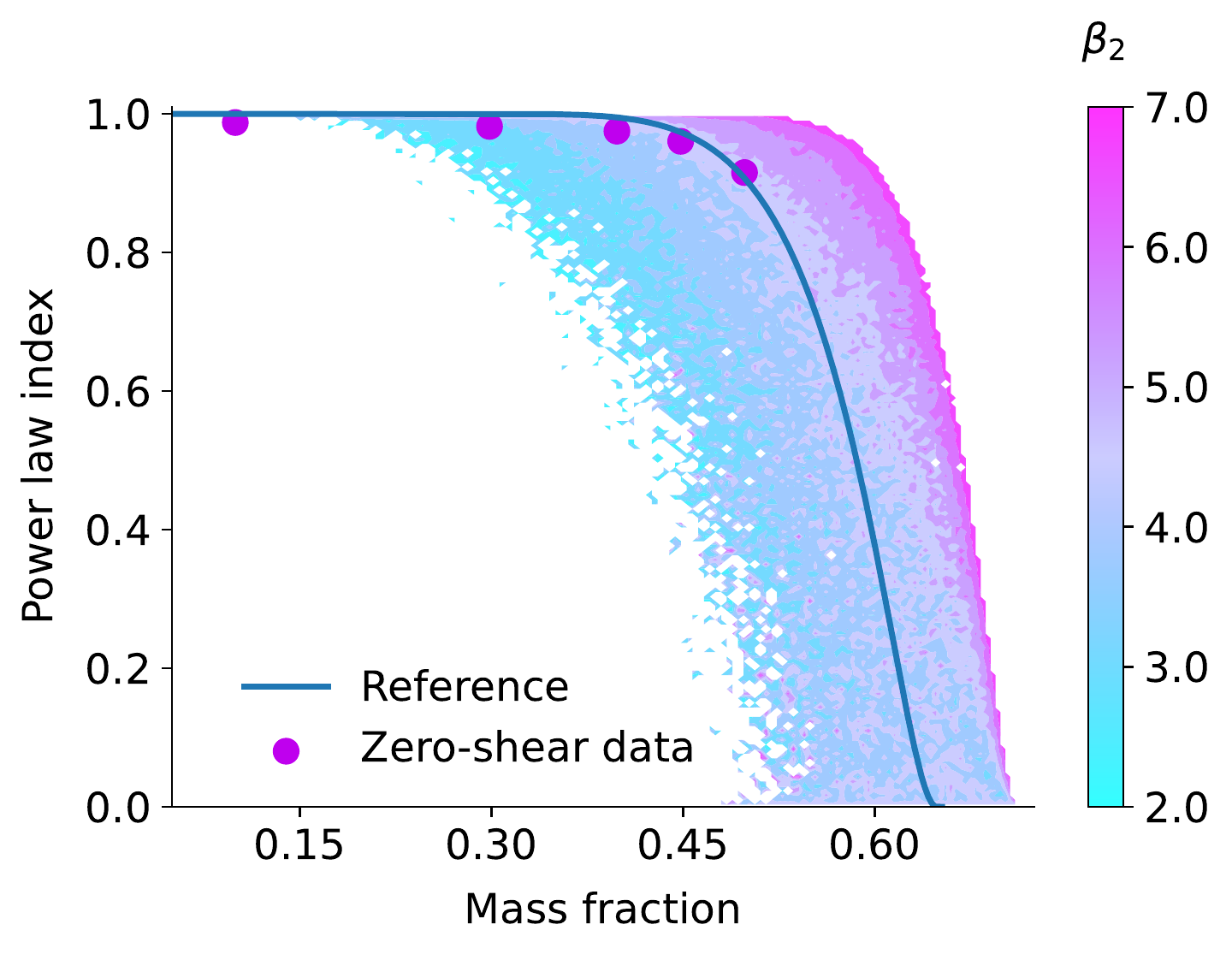}
	\end{subfigure}
	\begin{subfigure}{.45\textwidth}
		\caption{}
		\includegraphics[width=\textwidth]{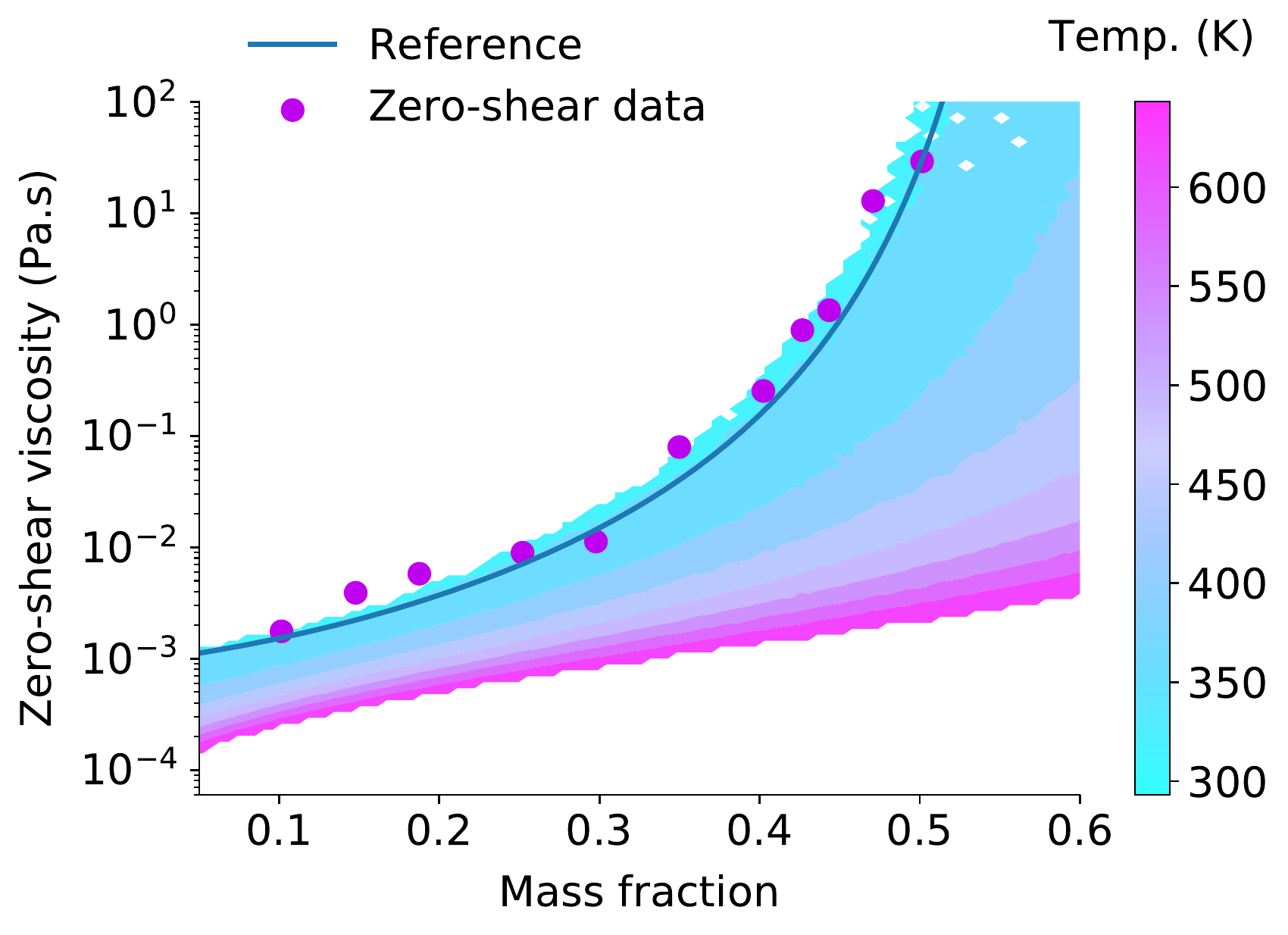}
	\end{subfigure}
	\caption{MC simulation results for the power-law index (a) and zero-shear viscosity (b) as a function of solids volume fraction (lignosulfonates case). The contours show power-law index and zero-shear viscosity dependence on $\beta_2$ and temperature, respectively. Experimental results and the model's prediction for the reference condition is also shown \citep{Vainio_2008}.}
	\label{fig:MC_idx-LS}
\end{figure}

\Cref{fig:MC_idx-LS,fig:MC_idx-BL} illustrates the power law index and zero-shear viscosity for BL and LS, respectively. These are a function of mass fraction on the $x$ axis with parameter $\beta_2$ or temperature as the contours, for the power law index or zero-shear viscosity, respectively. The distributions of $\mathcal{K}_{\beta_2}$ and $\mathcal{K}_{f_B}$ define how the these contours vary. Their limits were set based on a percentage difference from the reference values - blue line in \cref{fig:MC_idx-LS,fig:MC_idx-BL}. This curve is obtained by non-linear least squares regression with the data points, presented in \cref{fig:MC_idx-LS}, using \cref{eq:eta0,eq:n} for the zero-shear viscosity and power law index, respectively.

The parameter $\beta_2$ determines how sharply the power-law index decreases when macroscopic particle clusters form or when the critical fraction is reached. The curve then steadily decreases until reaching zero, with this concentration corresponding to the maximum packing fraction of the mixture. The overall shape of the power-law index curves for both mixtures is similar, with the BL curve starting to decrease at slightly higher mass fractions when compared to LS. The influence of $\beta_2$ on the BL power-law index contour is less pronounced than in the LS case. Also, $\beta_2$ values above the reference curve do not influence the point of maximum packing, while for LS,
this value varies with $\beta_2$. The main difference between the two mixtures relating to the power-law index is the lower molecular weight, on average, of LS compared to BL. Therefore, the $\beta_2$ parameter can only influence the maximum packing fraction at the low end of the molecular weight distribution.

The zero-shear viscosity prediction contours have a similar shape between mixtures, with the exponential behaviour more evident for LS. The model predictions fail to capture the experimental values at lower temperatures and concentrations,  seen in \cref{fig:MC_idx-BL}, where data for different temperature levels is available. Nonetheless, this should not significantly affect the accuracy of the thermophysical model, as the goal is to simulate high concentration, high-temperature mixtures. The higher values of zero-shear viscosity for LS for the same concentration should translate into higher overall viscosity than BL.

\begin{figure}
	\centering
	\begin{subfigure}{.45\textwidth}
		\caption{}
		\vspace{-1ex}
		\includegraphics[width=\textwidth]{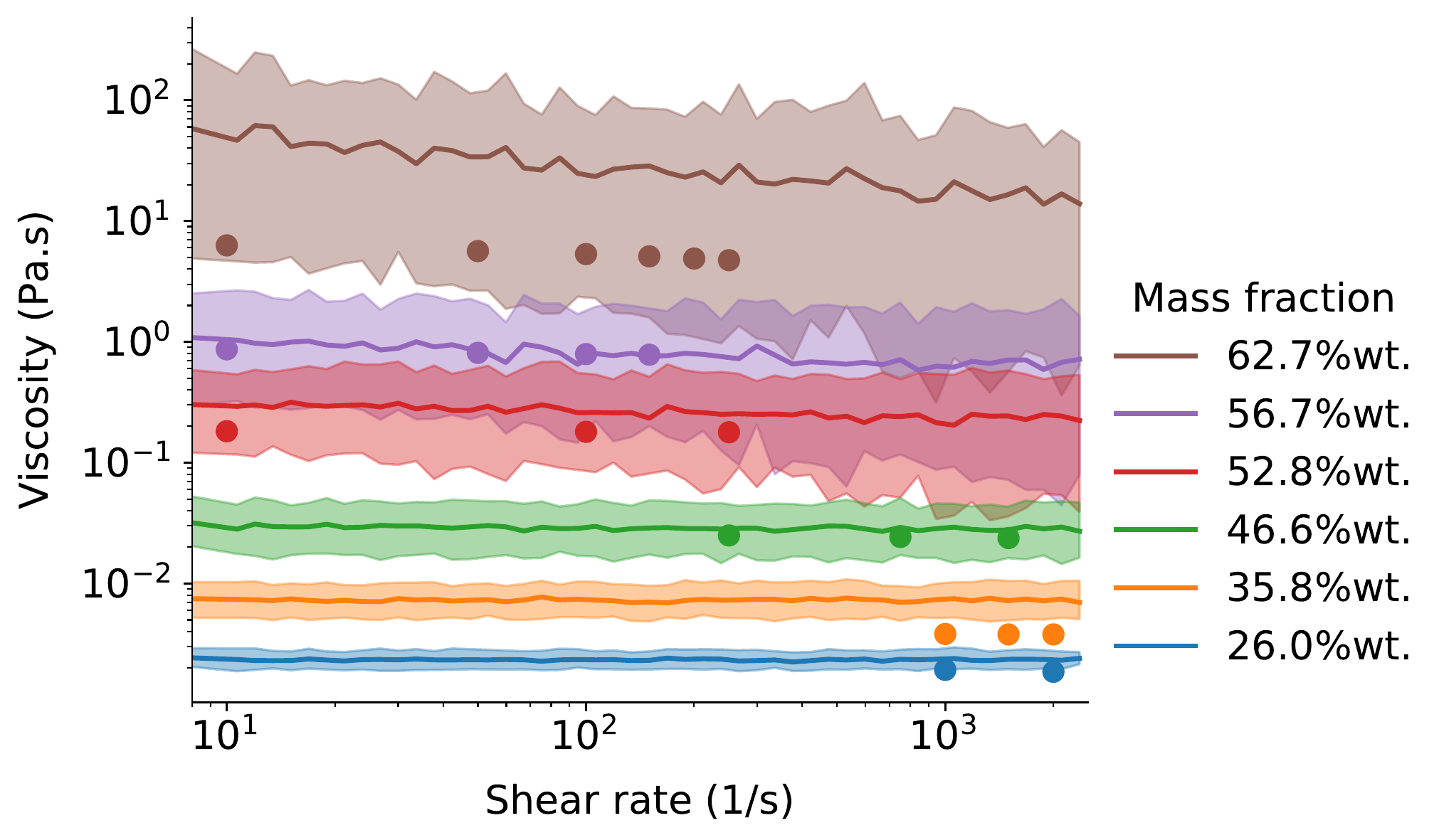}
	\end{subfigure}
	\begin{subfigure}{.45\textwidth}
		\caption{}
		\vspace{-1ex}
		\includegraphics[width=\textwidth]{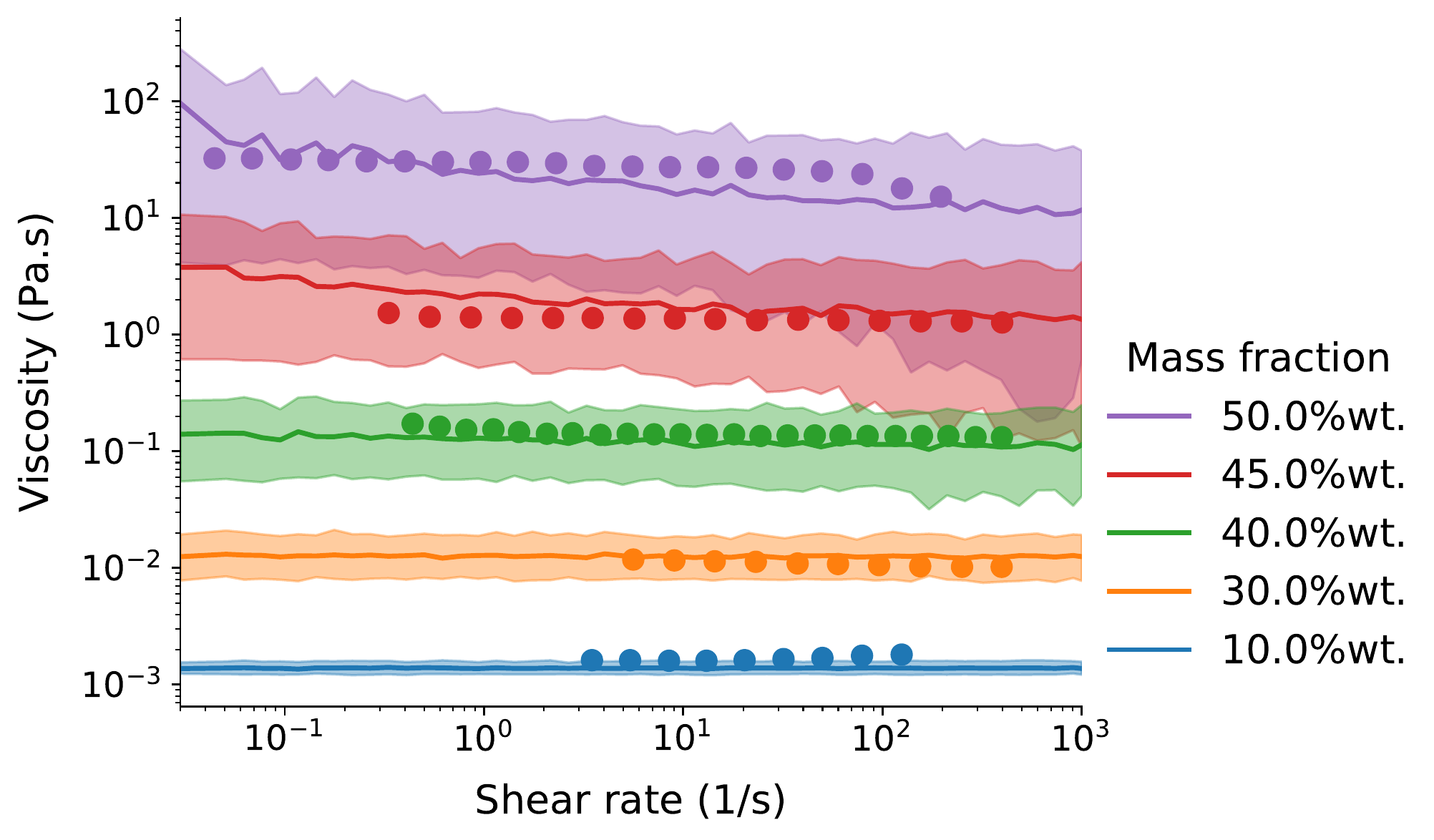}
	\end{subfigure}
	\caption{Mean viscosity and 90\% confidence interval as a function of shear rate for several solid concentrations. Black liquor results are denoted by (a) and lignosulfonate results by (b), at temperatures of $333.15K$ and $298.15 K$, respectively. Experimental values are shown for reference as circles.}
	\label{fig:visc_gamma}
\end{figure}

\Cref{fig:visc_gamma} show the resulting viscosity as a function of shear rate for BL and LS for selected solid concentrations, along with the respective confidence intervals. The model reasonably predicts viscosity for LS across all concentrations, failing at low shear rates. The model predictions for BL also deviate from the experimental points at low shear rates. However, the threshold value at which this deviation starts to happen is around two orders of magnitude higher than with LS. The low accuracy in this operation region requires special attention when performing simulations at low shear conditions.

\subsection{Temperature} \label{ssec:temp}

\begin{figure}
	\includegraphics[width=0.55\textwidth]{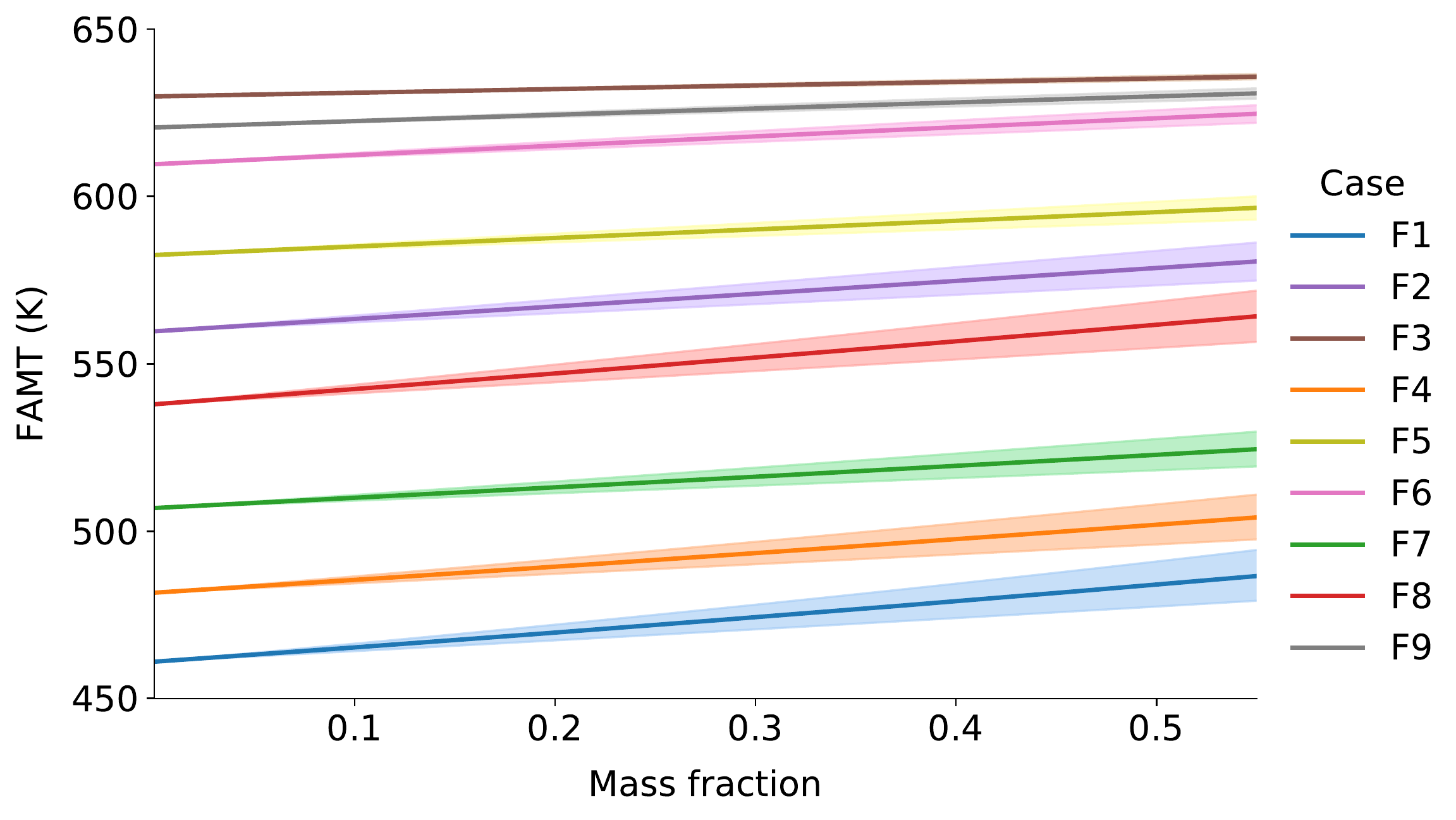}
	\caption{Frozen adiabatic mixing temperature for the different test case operating conditions.}
	\label{fig:FAMT}
\end{figure}

The frozen adiabatic mixing temperature (FAMT) is shown in \cref{fig:FAMT}, grouped in sets of operating conditions. Each sub-figure contains the three different temperature levels considered, differing in total flow rate and hot to cold flow (H/C) ratio. All FAMT curves show a linear relationship with mass fraction, and their respective confidence interval slightly increases with concentration. Hot stream temperature and the H/C ratio are the variables that influence FAMT the most. The total flow ratio and consequently the Reynolds number does not show a significant impact on FAMT. The confidence interval for cases at higher temperatures and flow ratios is narrow due to the relative differences in heat capacity between hot compressed water and lignin particles. The heat capacity of water increases exponentially at temperatures close to the critical point, while a linear polynomial describes lignin throughout the entire temperature range. Since there are no uncertain parameters in computing pure water properties, cases where the FAMT is close to the critical point or the H/C ratio is high will inherently lead to more accurate model predictions. The Sobol indices for FAMT are not presented as heat capacity is the sole contributor to its uncertainty.

Considering the results for the FAMT, the thermophysical influence on the energy balance is not very relevant even at high lignin loadings. The HTL process is restricted in the maximum  mixing temperature and H/C ratios employed. The pure water and aqueous mixtures must remain sub-critical as salt precipitation occurs near the critical point. Therefore, the upper limit of the FAMT confidence intervals must not be higher than the water's critical temperature. The H/C ratio, also an essential variable in controlling the reactor's temperature,  can reduce the economic feasibility of the HTL process, so it must not result in an excessively diluted mixture.

\subsection{Mixing time ratio} \label{ssec:mixing}

\begin{figure}
	\begin{subfigure}{.45\textwidth}
		\centering
		\qquad$\mathbf{573.15 K}$
		\includegraphics[width=\textwidth]{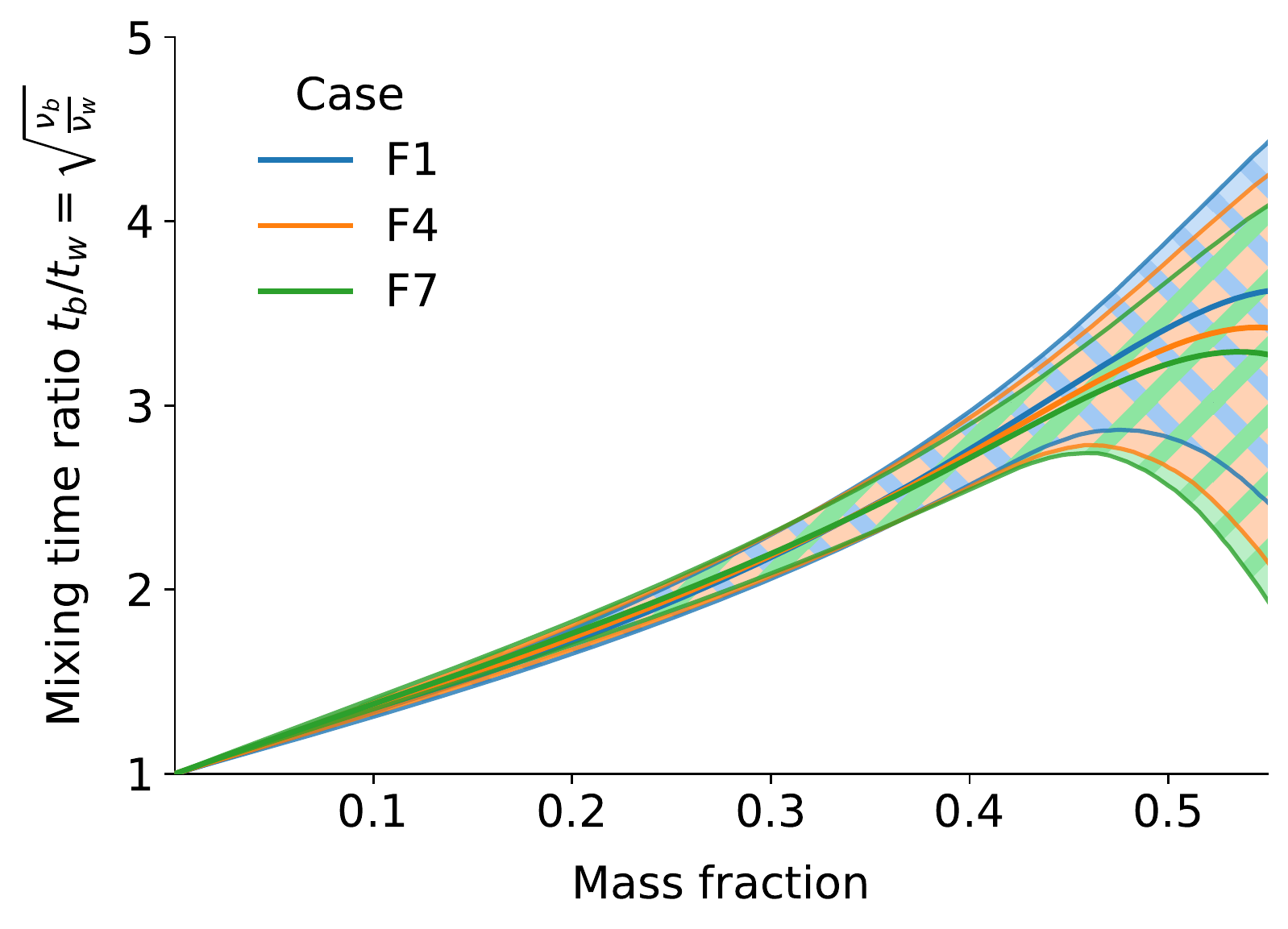}
		\includegraphics[width=\textwidth]{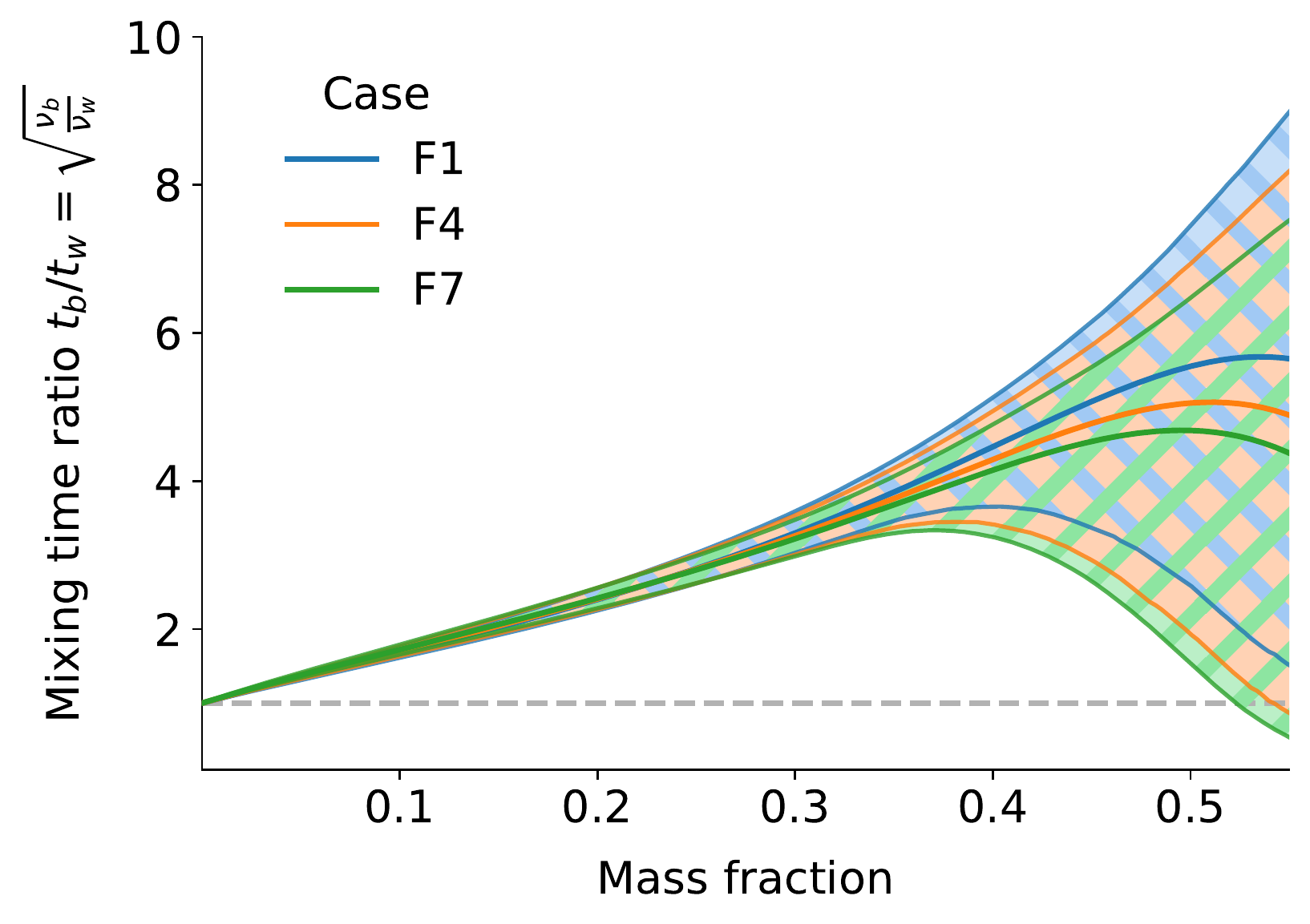}
	\end{subfigure}
	\begin{subfigure}{.45\textwidth}
		\centering
		\qquad$\mathbf{623.15 K}$
		\includegraphics[width=\textwidth]{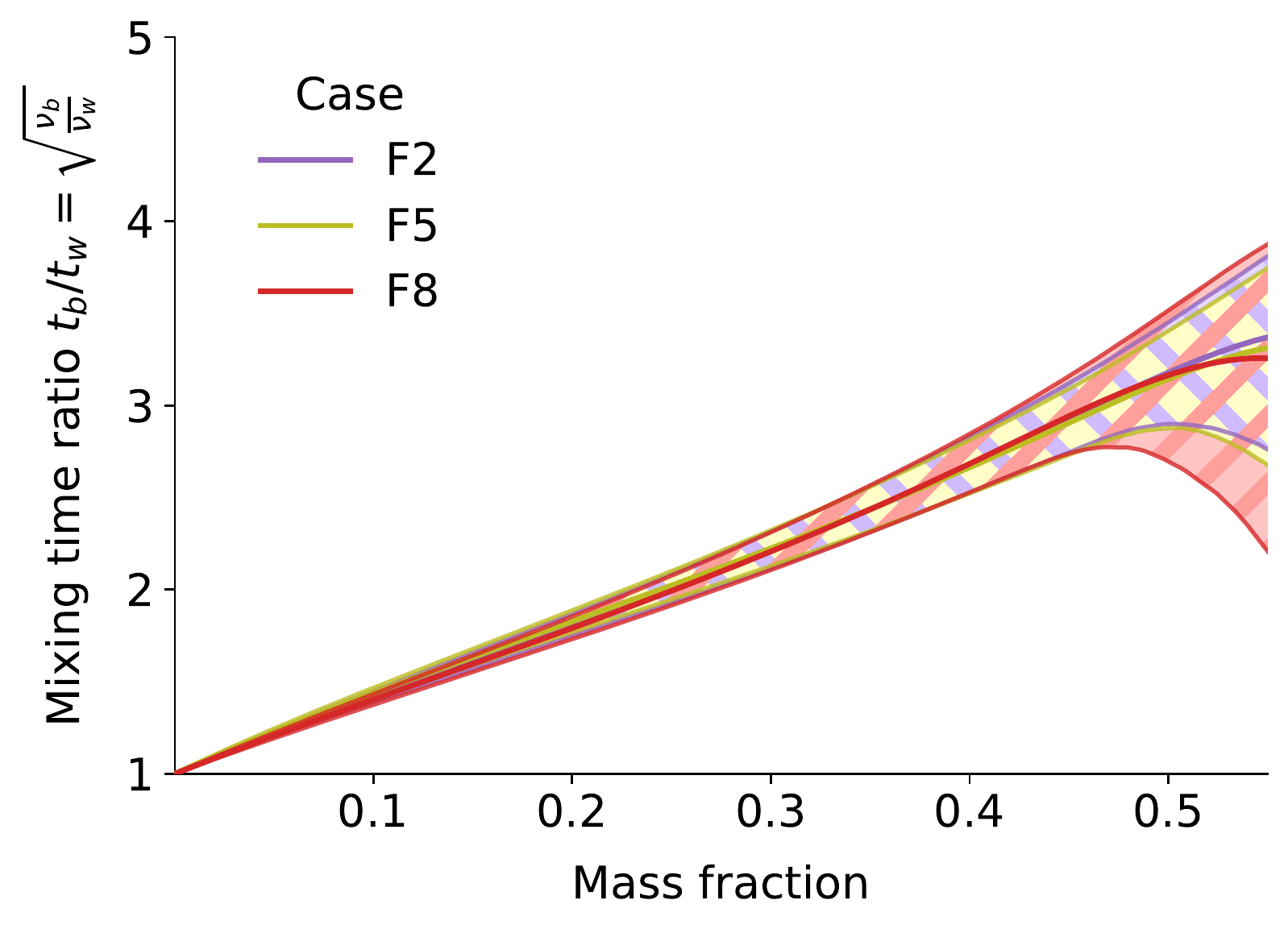}
		\includegraphics[width=\textwidth]{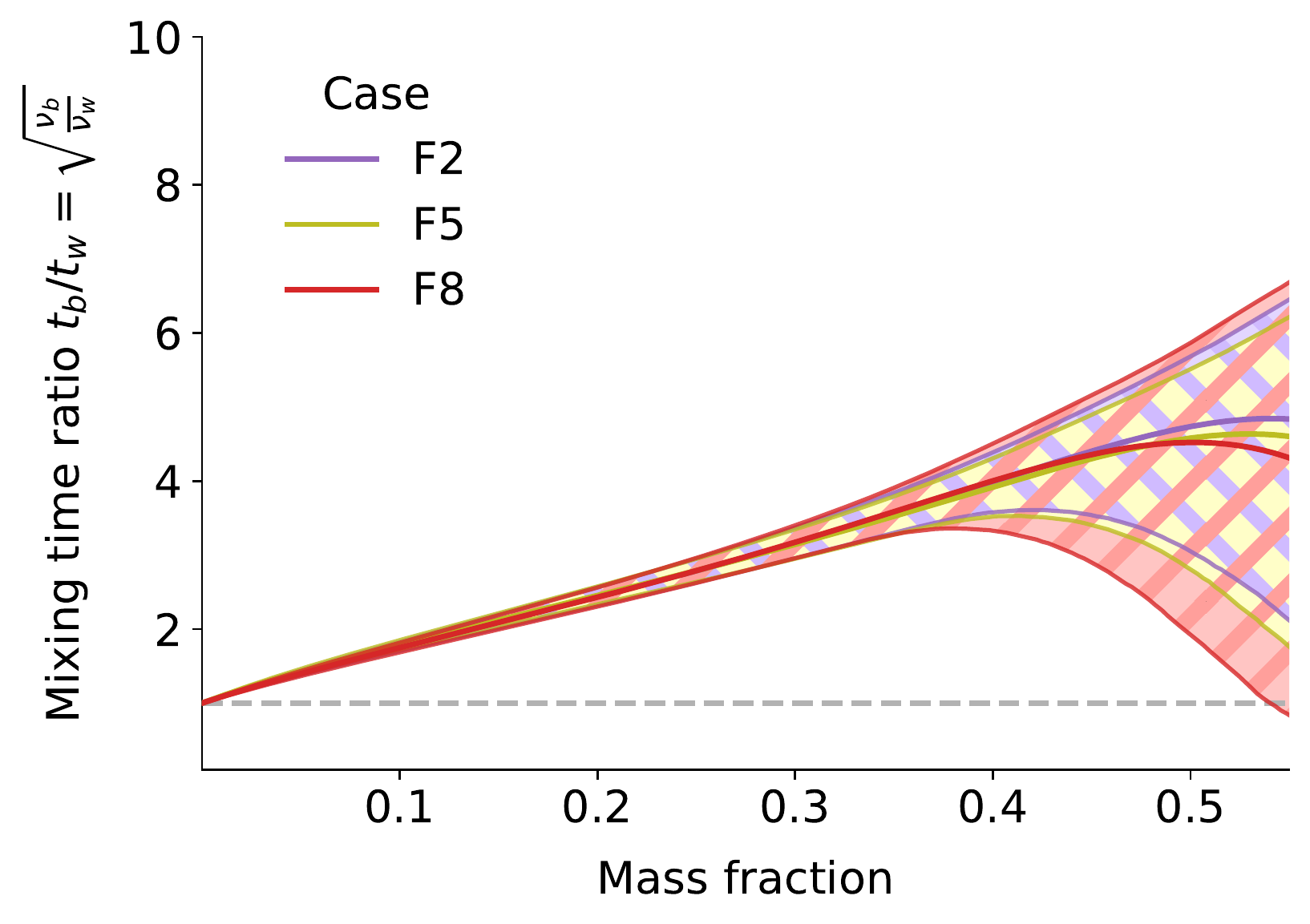}
	\end{subfigure}
	\begin{subfigure}{.45\textwidth}
		\centering
		\qquad$\mathbf{643.15 K}$
		\includegraphics[width=\textwidth]{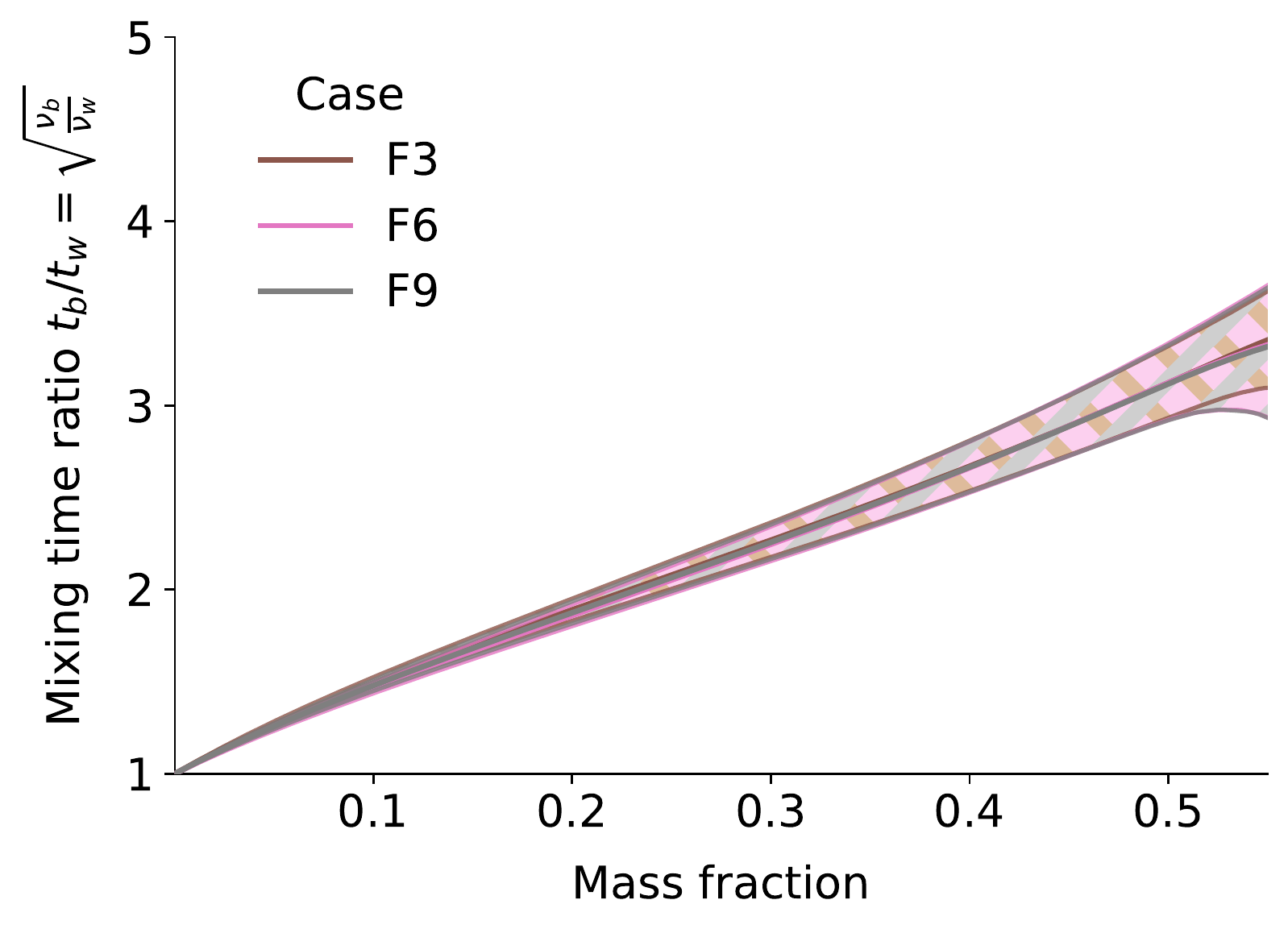}
		\includegraphics[width=\textwidth]{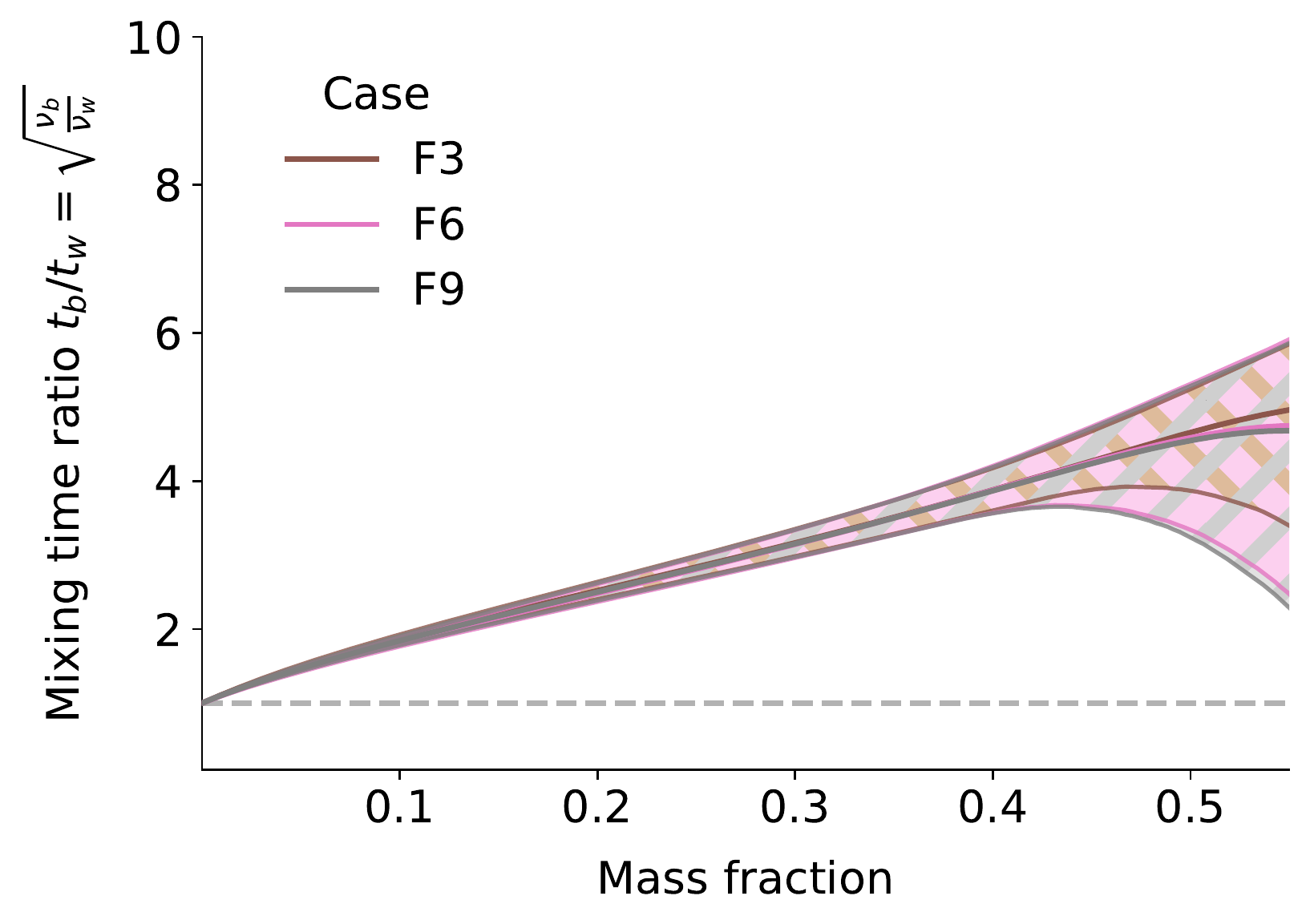}
	\end{subfigure}
	\caption{Mixing time for black liquor (top) and lignosulfonates (bottom).}
	\label{fig:mix_t}
\end{figure}

Mixing aqueous lignin mixtures with hot compressed water is crucial in the HTL process. The latter will provide the necessary reaction heat and act as a solvent for the dissolved lignin in the cold stream. Good levels of micromixing are required  to achieve this, and these are measured indirectly by the mixing time ratios. The lower the mixing time ratio, the easier it is to achieve good mixing at all scales since it is much easier to mix pure water streams. \Cref{fig:mix_t} shows the mixing time as a function of mass fraction, with higher temperatures and total flow rates contributing to a lower mixing time ratio and consequently a narrower confidence interval. As solid concentration increases, the mixing time confidence interval gets wider, particularly above mass fractions of $50\%$. The thermophysical model's low accuracy at high concentrations limits its applicability. However, it is still reasonably acceptable above $36.6\%_{wt.}$ and up to $45-50\%_{wt.}$, depending on process conditions, which is a concentration range that should ensure economic feasibility to the HTL process \citep{Knorr_2013}.

The viscosity consistency index is modelled as a weighted average of water and zero-shear aqueous lignin viscosity, resulting in the latter having a much more significant contribution to the final viscosity value at low temperatures and H/C ratios. This difference in magnitude is lower at high temperatures, close to the water's critical point and high H/C ratios due to the diluted mixture exiting the reactor. Cases with higher temperatures show smaller confidence intervals due to the differences in the relative contribution of the water and biomass components to the final viscosity value.

\begin{figure}
	\begin{subfigure}{.435\textwidth}
		\centering
		\caption{}
		\includegraphics[width=\textwidth]{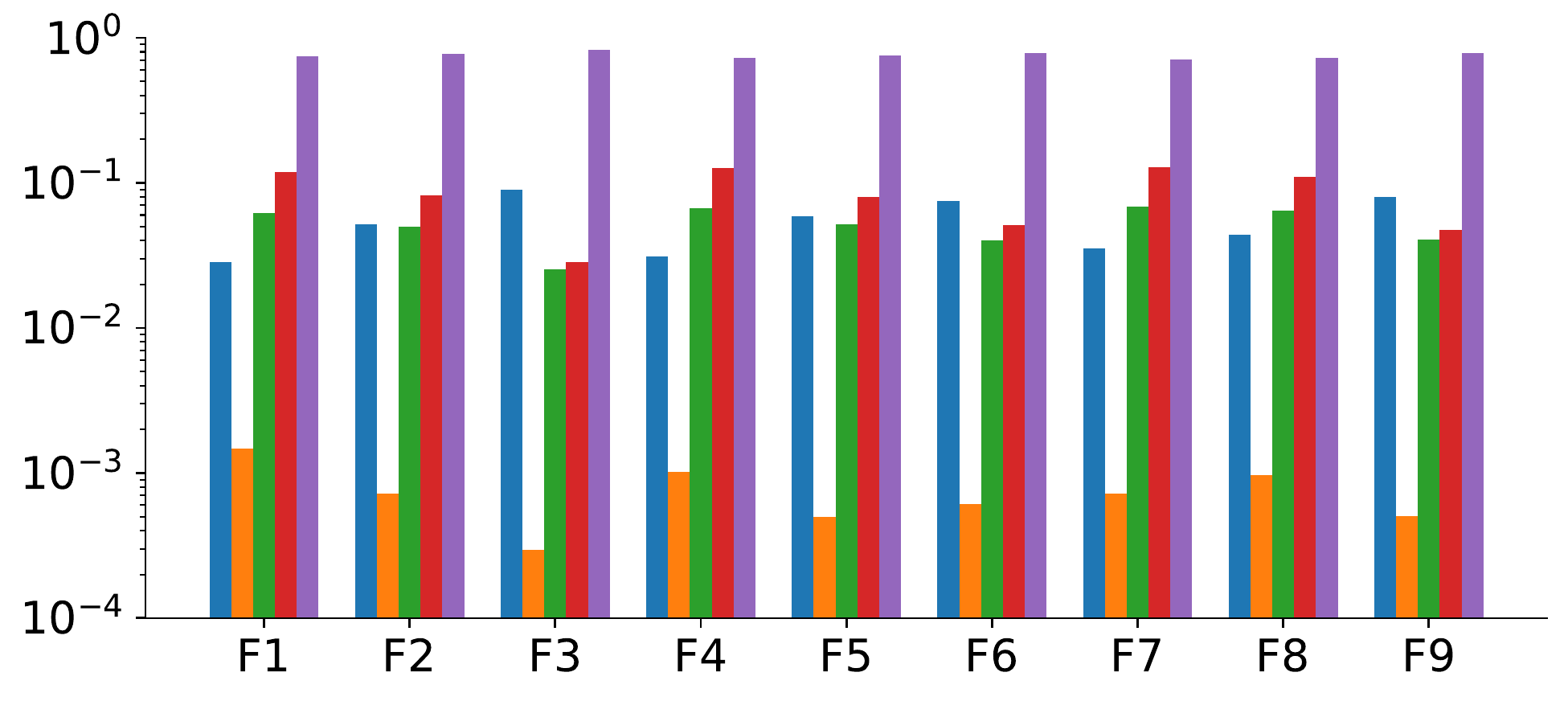}
	\end{subfigure}
	\begin{subfigure}{.5\textwidth}
		\centering
		\caption{}
		\includegraphics[width=\textwidth]{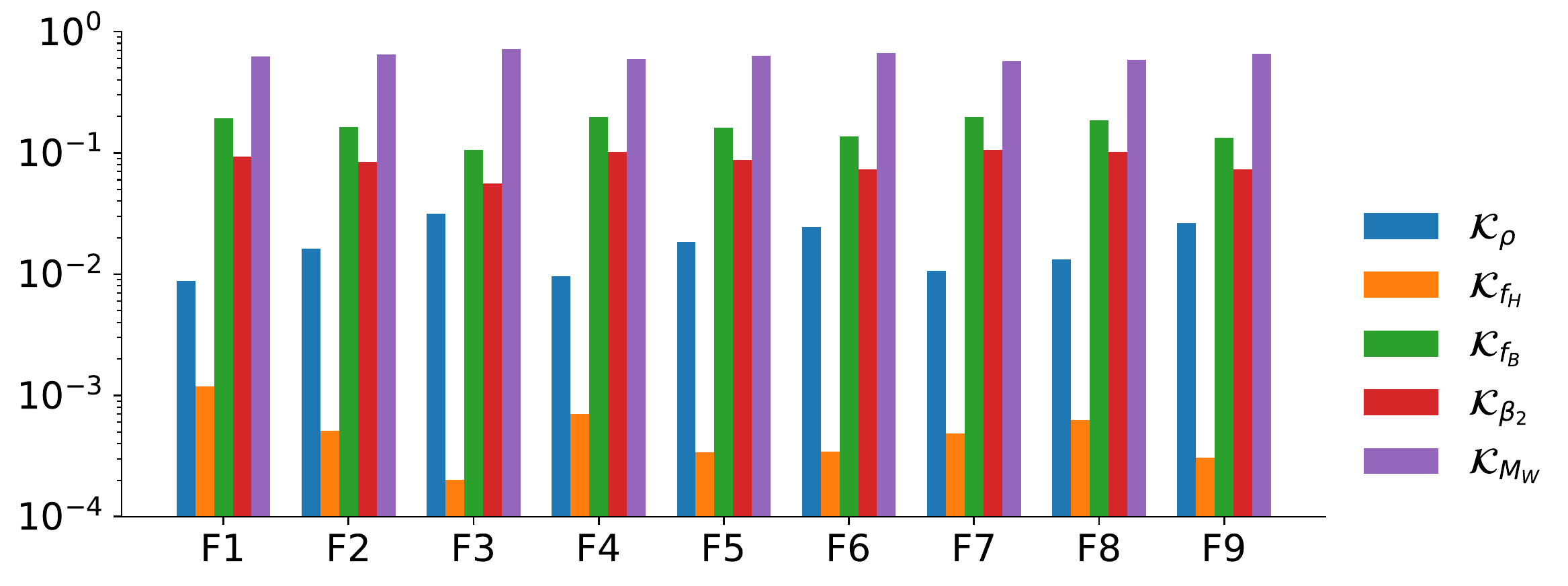}
	\end{subfigure}
	\caption{Mixing time ratio first order average Sobol indices for each test case. Black liquor cases are denoted by (a) and lignosulfonate cases indicated by (b).}
	\label{fig:sobol_mix}
\end{figure}

The differences in mixing time between BL and LS are substantial and show that the latter is considerably more sensitive to the same degree of uncertainty when compared to BL. The overall shape and behaviour of the curves are similar between the two mixtures for all tested cases, the only difference being the magnitude of the confidence interval. This could be due to the relatively higher contribution of $\mathcal{K}_{f_B}$ to the mixing time ratio uncertainty for the LS mixture, as can be seen in \cref{fig:sobol_mix}. Additionally, \cref{fig:sobol_mix} reveals that $\mathcal{K}_{M_W}$ has the highest contribution to the mixing time ratio uncertainty for both mixtures, likely due to the wide  value range of the molecular weight distributions.
 
Some test cases show a lower limit to the mixing time ratio confidence interval below unity, which can be an artefact of the UQ procedure, as the physical nature of the variables is not considered. Therefore, the percentiles computation is not restricted and can even take negative values, which is impossible for any studied QoIs. Additionally, the uncertainty introduced by the parameters $\mathcal{K}_{M_W}$, $\mathcal{K}_{\beta_2}$ and $\mathcal{K}_{f_B}$ combined with high solids mass fractions and exceptionally high Reynolds numbers can lead to shearing conditions where the mixing time ratio can take values below unity. For the same temperature, cases with higher Reynolds numbers and consequently shear rates (cases F7-9) always have lower $5\%$ percentile values than the other cases due to the increased likelihood of the mixing time ratio being below unity. The latter explains the inflexion point and trend change in the confidence interval lower limit at high mass fractions.

The micromixing time between the aqueous lignin mixture and hot compressed water can increase up to ten-fold for LS and up to five-fold for BL compared to a scenario with two streams of pure water. Increasing temperature, the H/C ratio and total flow rate all aid in reducing the mixing time. However, practical considerations limit how far this QoI can be reduced without introducing changes to the process. Due to high pressures and viscous flows in HTL processing, the pumping requirements are considerable, and both high H/C ratios and total flow rates will further increase these requirements. Given the high variability of lignin mixture properties, the counter-current mixing alone may not mix the two streams at the molecular level. In this situation, additional mechanical energy in the form of an impeller is likely necessary.

\subsection{Residence time distributions} \label{ssec:RTD}

\begin{figure}
	\centering
	\includegraphics[width=0.45\textwidth]{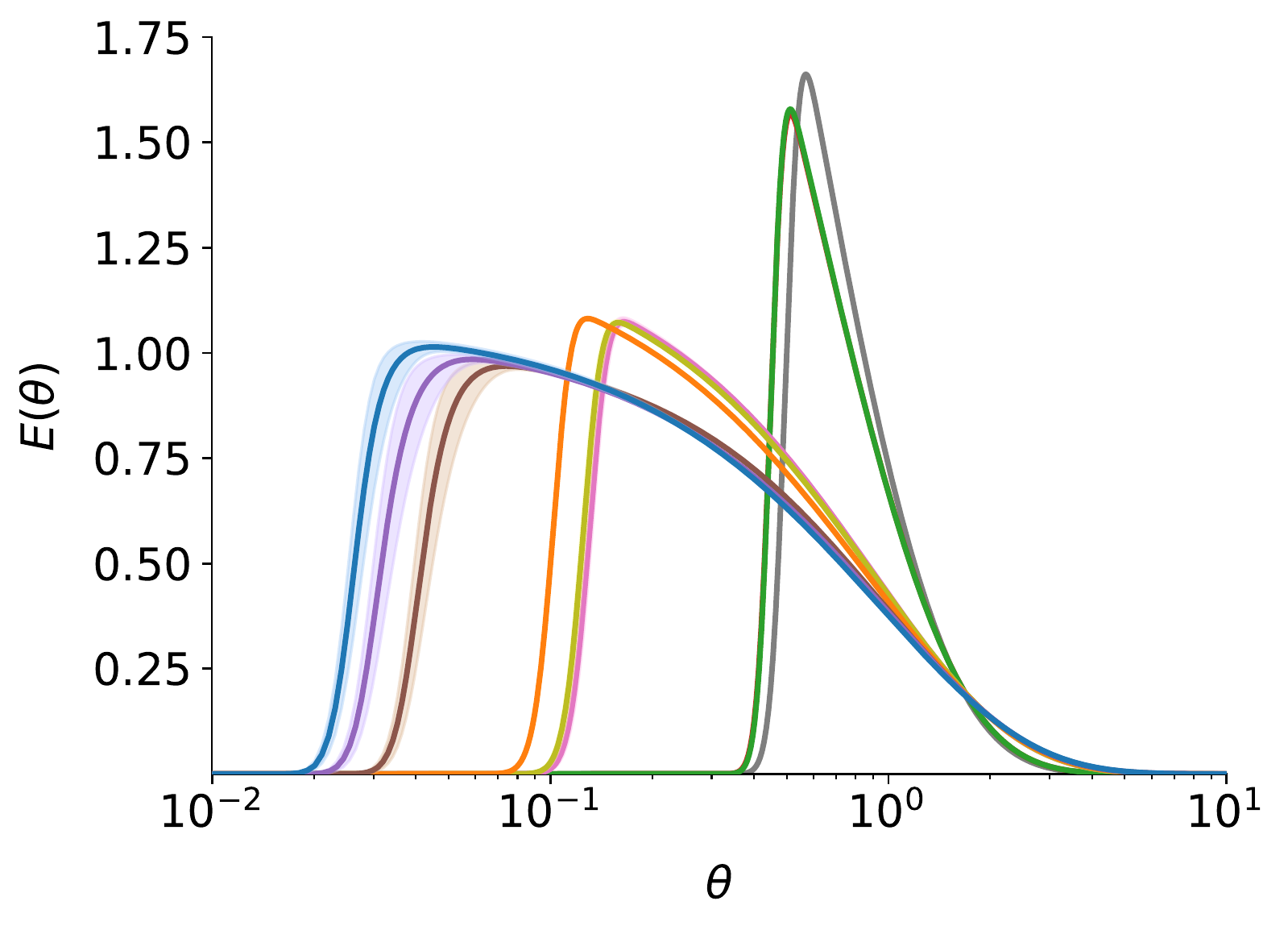}
	\includegraphics[width=0.45\textwidth]{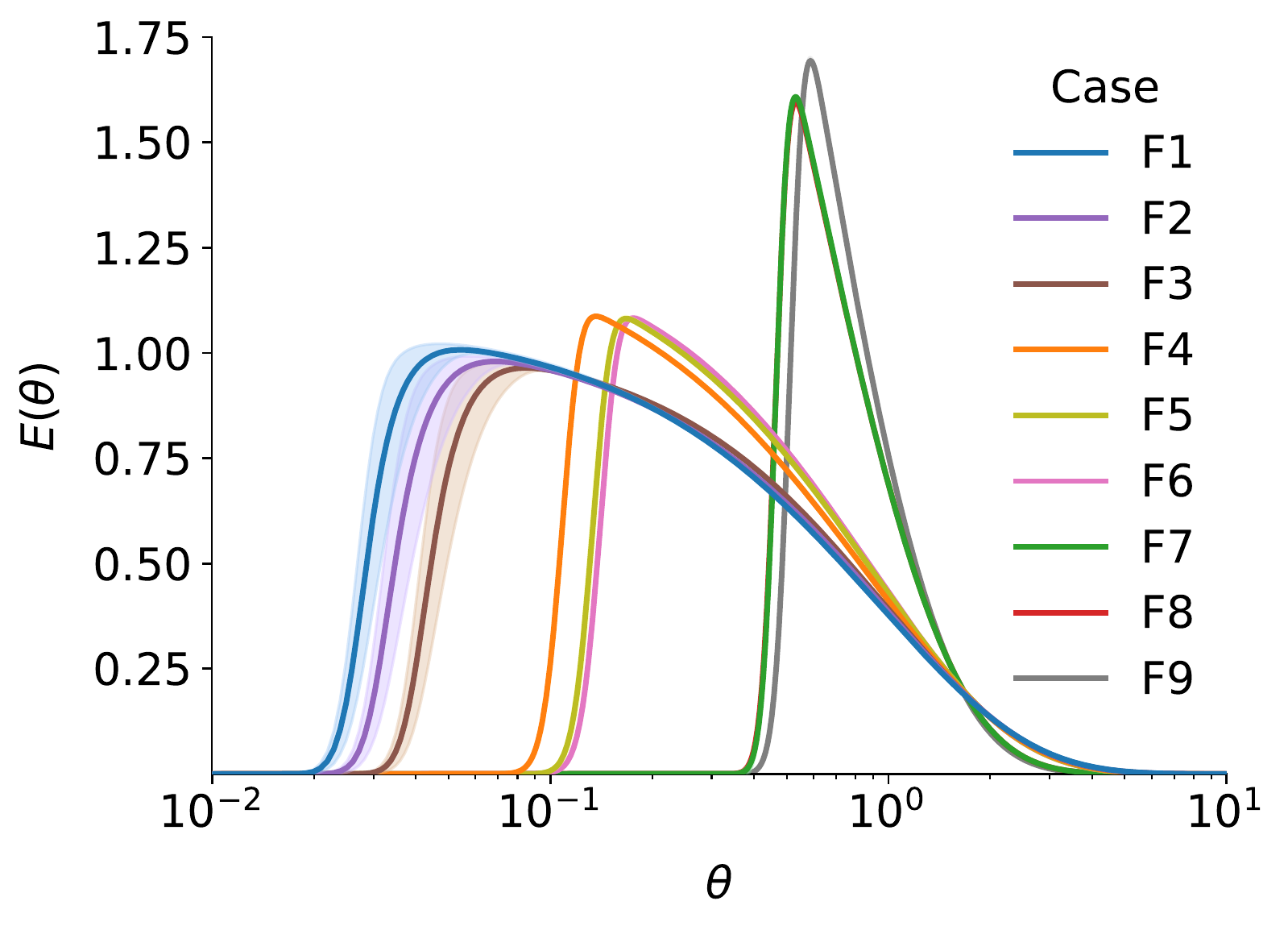}
	\caption{Non-dimensional residence time distribution curves for the studied test cases. Left shows lignosulfonates, right shows black liquor. $\theta$ and $E(\theta)$ are the non-dimensional time and tracer concentration, respectively.}
	\label{fig:rtd_UQ}
\end{figure}

\Cref{fig:rtd_UQ} shows the RTD curves for the considered test cases. There is a clear distinction between cases F1-3, F4-6 and F7-9 that correspond to the three different levels of total flow rate considered. Thermophysical property variations only impact the RTD curves at low $\theta$ values and laminar flow regimes. At these conditions, the fluid's thermophysical properties are dominated by molecular effects, which have an appreciable impact on the RTD curve, translating into the confidence intervals for cases F1-3 seen in \cref{fig:rtd_UQ}. However, due to the intrinsically small scale of this phenomenon, it is only relevant at low mass flow rates.

An RTD is a measurement of macroscopic mixing, tied to large scale motions and turbulent diffusion. The RTD curves show virtually no differences between the BL and LS, consistent with what was previously said about the thermophysical properties influence on this QoI. The results show that the reactor is dynamically identical for the two fluids, however, the RTD cannot provide information related to the moment when mixing occurs or whether the outlet stream is fully mixed at the microscale \citep{Baldyga_1984}. The  latter is given instead by the mixing time ratio, discussed before in \cref{ssec:mixing}. Therefore, according to this model, BL and LS have identical macromixing behaviour and only differ at the microscale, with LS requiring more energy to achieve good mixing.

\begin{figure}
	\centering
	\includegraphics[width=.45\textwidth]{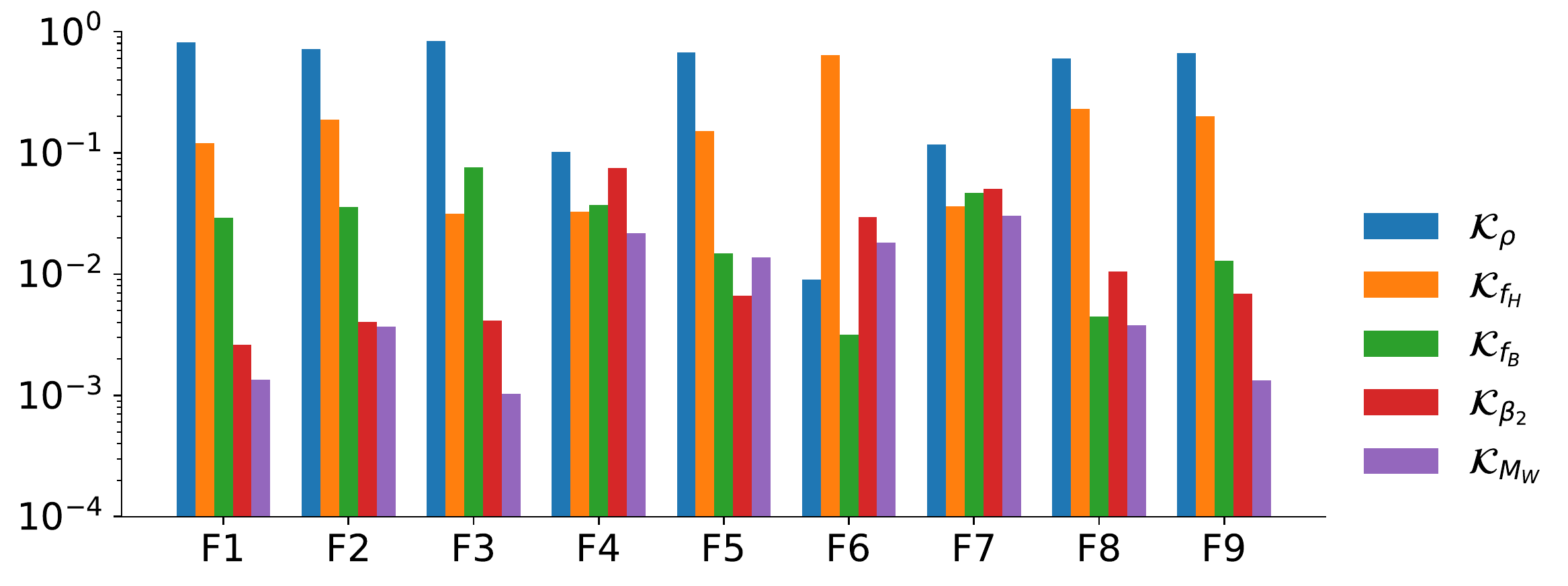}
	\caption{Residence time distribution first order average Sobol indices for each case.}
	\label{fig:sobol_rtd}
\end{figure}

The Sobol indices, presented in \cref{fig:sobol_rtd}, show that viscosity has a minimal impact on the RTD curves variability. Additionally, cases F4 and F7 show very drastic changes in the contribution of density to the overall RTD curve uncertainty. These are likely numerical errors due to the already negligible contribution of the thermophysical properties to the RTD uncertainty. Cases F1-3 support this by showing a consistent contribution of the parameters $\mathcal{K}_{\rho}$, followed by $\mathcal{K}_{f_H}$ to the results' uncertainty. Since these cases have a laminar flow regime, the influence of the fluid's properties on the RTD curve uncertainty is not so low to cause numerical errors.

The shape and magnitude of the RTD curves start to become dominated by flow advection with increased Reynolds number. Cases F4-6 show a $E(\theta)$ maximum slightly higher than cases F1-3, and the respective $\theta$ values are closer to unity. There are no discernable confidence intervals, meaning the fluid's properties have virtually no effect on the RTD curves. Cases F7-9 present curves with shorter tails and a significantly higher maximum, located at $\theta$ values much closer to unity than all other cases. Therefore, most fluid particles will exit the reactor at the mean residence time,  minimising dead zones and flow by-passes and ensuring a similar processing experience. Consequently, the primary HTL reactions are promoted, and unwanted secondary reactions are less likely to happen. \Citet{VanGerven_2009} identify this as one fundamental principle of process intensification, necessary to ``deliver ideally uniform products with minimum waste".

\section{Conclusions} \label{sec:conclusions}

The present study aimed to develop a thermophysical model representing two aqueous lignin mixtures, black liquor and lignosulfonates, at high solid concentrations and temperatures and pressures common to hydrothermal processes. This model establishes a simulation framework that does not  assume pure water properties for the biomass carrying stream. Thus, the computational studies performed with this model are not limited to diluted biomass mixtures. The second aim of this study was to combine a novel UQ procedure based on polynomial chaos expansions with the reactor simulations and respective quantities of interest. Due to the heterogeneous nature of lignin mixtures, several thermophysical model parameters may have different values. The uncertainty quantification procedure allows these uncertain parameters to assume values drawn from a probability distribution function instead of a single value, as is customary in deterministic simulations. The influence of the thermophysical model uncertain parameters on the simulation quantities of interest can thus be assessed and a confidence interval determined.

The analysis showed that the fluid's thermophysical properties influence on FAMT only becomes relevant at temperatures significantly lower than the water's  critical point due to the relative difference in lignin and water heat capacities. The latter increases by a factor of four at temperatures near the critical point, having a much more significant contribution to the final heat capacity value than lignin, which is modelled by a linear polynomial. The influence of thermophysical properties on the residence time distribution curves are only evident at laminar flow regimes, where molecular effects become relevant compared to flow variables.
The mixing time ratio results show a much more considerable influence of the mixture's properties than the other two quantities of interest, with the accuracy of the results dropping severely above concentrations of 50\%. For the same concentration, the micromixing time can increase between two and a half to five-fold and two to ten-fold for black liquor and lignosulfonate mixtures, respectively, when compared to mixing two pure water streams. Any increase in micromixing time will translate into higher energy costs to achieve the same level of mixing. Given the high uncertainty of these results, it is unlikely that counter-current mixing alone can adequately mix high concentration lignin mixtures with hot compressed water without additional mechanical energy, in the form of an impeller, for example.

Since the scope of this study was the thermophysical model development, the reactor simulations are highly simplified. The most significant limitations are the one-dimensional model, which does not include radial dispersion or compute heating rates and the absence of a turbulence model. Despite this, the mean values for the computed quantities of interest should remain the same. A more detailed flow model mainly influences the shape and magnitude of the quantities of interest confidence intervals. While the uncertainty quantification procedure mitigated the limited data issues, the accuracy of the model predictions ultimately depends on the validity of its assumptions. Measurements and experimental data can only confirm these.

The thermophysical model and uncertainty quantification procedure can be extended to more rigorous simulations, such as computational fluid dynamics, without changing model parameters. A more rigorous flow model can compute additional quantities of interest, and aqueous lignin's non-Newtonian behaviour influence on turbulence can be studied, which is particularly relevant for transitional flow regimes.

%% References
\bibliographystyle{elsarticle-num-names}
\bibliography{biblio.bib}

\section*{Declaration of Competing Interest}
The authors declare that they have no known competing financial interests or personal relationships that could have influenced the work reported in this paper.

\section*{Acknowledgements}
This work is part of project 70442681, financed by the Department of Energy and Process Engineering, Faculty of Engineering Science, Norwegian University of Science and Technology.

\end{document}